# Water ice in the dark dune spots of Richardson crater on Mars

Kereszturi A.[1], Vincendon M.[2], Schmidt F.[3]

[1] Collegium Budapest, Institute for Advanced Study, H-1014 Budapest, Szentháromsag 2. Hungary

[2] Department of Geological Sciences, Brown University, Providence, Rhode Island, USA

[3] IDES, Université Paris-Sud 11; CNRS/INSU, Bâtiment 509, Université Paris-Sud, 91405 Orsay, France

## Abstract

Interfacial liquid water has been hypothesized to form during the seasonal evolution of the dark dune spots observed in the high latitudes of Mars. In this study we assess the presence, nature and properties of ices – in particular water ice – that occur within these spots using HIRISE and CRISM observations, as well as the LMD Global Climate Model. Our studies focus on Richardson crater (72°S, 179°E) and cover southern spring and summer ($L_S$ 175° - 341°). Three units have been identified of these spots: dark core, gray ring and bright halo. Each unit show characteristic changes as the season progress. In winter, the whole area is covered by $CO_2$ ice with $H_2O$ ice contamination. Dark spots form during late winter and early spring. During spring, the dark spots are located in a 10 cm thick depression compared to the surrounding bright ice-rich layer. They are spectrally characterized by weak $CO_2$ ice signatures that probably result from spatial mixing of $CO_2$ ice rich and ice free regions within pixels, and from mixing of surface signatures due to aerosols scattering. The bright halo shaped by winds shows stronger $CO_2$ absorptions than the average ice covered terrain, which is consistent with a formation process involving $CO_2$ re-condensation. According to spectral, morphological and modeling considerations, the gray ring is composed of a thin layer of a few tens of μm of water ice. Two sources/processes could participate to the enrichment of water ice in the gray ring unit: (i) water ice condensation at the surface in early fall (prior to the condensation of a $CO_2$ rich winter layer) or during winter time (due to cold trapping of the $CO_2$ layer); (ii) ejection of dust grains surrounded by water ice by the geyser activity responsible for the dark spot. In any case, water ice remains longer in the gray ring unit after the complete sublimation of the $CO_2$. Finally, we also looked for liquid water in the near-IR CRISM spectra using linear unmixing modeling but found no conclusive evidence for it.

Corresponding author's email: akos@colbud.hu , tel.: 36-06-30-3437-876. fax: 36-1- 240-7708. Regular postal address: Collegium Budapest Institute for Advanced Study, H-1011 Budapest, Szentharomsag u. 2. Hungary

## 1. Introduction

The aim of this work is to study the presence, origins and properties of water ice in the dark spots observed at the surface of the polar regions of Mars during local spring. These spots can give rise to flow-like features (Horvath et al. 2001, 2009). Assessing the presence and properties of water ice there is an important step while studying the possible contribution of liquid water to the formation of these features. The polar regions of Mars undergo a strong and rapid warming during spring (Kieffer and Titus, 2001). Local dark, sunlight absorbing environments such as the dark spots could be favorable environments for the formation of ephemeral liquid water on Mars if water ice is present at those locations during spring. It is already known that water ice is present at the kilometer scale in the seasonal and perennial ice caps of both hemispheres (Kieffer et al., 1976; Titus et al. 2003, Bibring et al., 2004). In the northern hemisphere, a water-ice ring follows the receding edge of the shrinking $CO_2$ ice cap (Kieffer and Titus, 2001; Bibring et al., 2005; Schmitt et al., 2005; Schmitt et al., 2006), partly because it appears from below the sublimated carbon dioxide seasonal cap and partly because it recondenses onto the perimeter of the remaining cold carbon dioxide ice (Wagstaff et al., 2008). On the contrary, no similar water ice annulus was observed in the southern hemisphere: water ice there consist of smaller water ice patches (Titus 2003, Bibring 2004, Titus, 2005; Langevin et al., 2006; Titus, 2008) which appear at the perimeter of the receding carbon dioxide cap, some of which remaining tens of kilometers away from the bright permanent $CO_2$ cap in summertime (Bibring et al., 2004). The high resolution imaging and spectroscopic dataset released by recent missions to Mars widen our ability to detect and study water in small, meter scaled environments such as the dark spots.

.

These Dark Dune Spots, which we abbreviate as DDS (Horváth et al., 2001; Gánti et al., 2003; Pócs et al., 2003), are part of the great number of strange seasonal features that appear during local spring in the polar regions of Mars, such as spiders (Kieffer, 2000; Kieffer et al., 2000; Ness and Orme, 2002; Piqueux et al., 2003; Christensen et al., 2005), dark spots (Malin et al., 1998, Malin and Edgett, 2001; Bridges et al., 2001; Zuber, 2003), flow-like features (Kieffer, 2003), Dalmatian spots (Kossacki and Kopystynski, 2004). Many of them could be the result of $CO_2$ gas outburst, but possibly other processes may also contribute like mass wasting (Hansen et al. 2010) or even flow of some liquid material (Möhlmann, Kereszturi 2010). These phenomena potentially affect the $CO_2$ sublimation and may influence the climate at regional scale (Schmidt et al., 2009, Schmidt et al., 2010). We will present new observational evidence related to one of those phenomena, the Dark Dune Spots, to further test and constrain the different proposed scenarios of formation for those features.

Recent observations and theoretical computations have suggested that ephemeral liquid water or brines could be present on Mars. Based on theoretical calculations, above around 180 K thin interfacial water layer (Möhlmann 2004) and possible solid-state greenhouse warming may produce bulk brines (Möhlmann 2010). At the polar landing site of Phoenix, not only solid ice was observed, but also spherical droplet-like features, slowly changing with time on the leg of the lander, that may be composed of dense salty solutions (Renno et al. 2009a, b). Ephemeral flow-like features emanate from Dark Dune Spots (DDSs) that could be formed by flowing brines or thin water layer lubricated grains have been recently studied in the northern and southern polar regions (Kereszturi et al. 2010a, Reiss et al. 2010) – although other interpretations, like dry mass wasting and avalanches could also be possible (Hansen et al.

2010). These flow-like features based on their appearance, dark color, and elevated temperatures are good candidates for the formation of liquid water.

Here we analyze dark spot-like features on the wintertime formed frost layer, which are called DDSs and which can give rise to flow-like features (Horvath et al. 2001, 2009), located on the Richardson crater (72°S, 179°E). We study these features using CRISM (Compact Reconnaissance Imaging Spectrometer for Mars) and HiRISE data to argue whether water – ice or liquid – is present in some part of these spots. $CO_2$ ice, that form in the polar regions and may cold trap $H_2O$ molecules and enhances their concentration, will also be studied in details. The high imaging performances of HiRISE (High Resolution Imaging Science Experiment, maximal spatial resolution 30 cm) are combined with the water detection capabilities of the near-IR CRISM experiment. Observational constraints about the nature and distribution of ice are compared to climate modeling predictions. A scenario is then proposed to explain the origin and evolution of water ice in this region.

## 2. Methods

This section contains the background information about the data acquired by the CRISM and HiRISE cameras onboard Mars Reconnaissance Orbiter, as well as a description of the modeling methods used to interpret them.

### 2.1 Observations

We analyzed CRISM spectral data (Murchie et al. 2007), with CAT-ENVI software (Morgan et al. 2009) in Richardson crater (72°S 179°E) using CRISM “FRT” observations, characterized by a high spatial and spectral resolution (18 m and 7 nm respectively). Near-IR wavelengths that range between 1 and 4 µm were considered, and observations were corrected for photometry and atmospheric gas absorptions (McGiure et al. 2009). A filtering method (Parente 2008) is used to reduce noise. The interpretation of spectra measured by push-broom sensors may be influenced by the effect called “spectral smile”: the mean wavelength of spectels and the spectral resolution can slightly change from one spatial pixel to the next (Ceamanos, Douté 2009). We do not correct for the spectral smile because we take spectra near the center of the CRISM image where band shifts are lower than 0.002 µm (Murchie et al. 2007), which has a minor impact on large absorption bands. The list of CRISM observations analyzed in this study is indicated in Table 1.

The CRISM spectral data were complemented by the analysis of the higher spatial resolution HiRISE images. We selected the highest resolution images acquired under proper illumination conditions to study the thickness of surface features using shadows (see Table 2.). We estimated absolute albedo values from HiRISE images, using the DN values of certain pixels from the RED channel images. We calculated the reflectivity with the formula I/F=(DN*SCALING_FACTOR)+OFFSET, and divided it with cos i, which gives an approach of lambertian albedo.

The study of the spatial and temporal distribution of $H_2O$ and $CO_2$ ice is performed using band depth values, representing the absorption efficiency at certain wavelength region, characteristic for certain molecules and minerals. The following parameters were used to analyze ices: “BD1500” (1.5 micron $H_2O$ ice band depth), “BD 1435” (1.435 micron $CO_2$ ice band depth), and “ICER1” ($CO_2$ and $H_2O$ ice band depth ratio) defined by Pelkey et al. 2007 (for detailed description of the calculation formulae see for the reference) on the basis of the parameters developed by Langevin et al. 2007. Very thin, micrometer thick layers of $H_2O$ ice can be detected in near-IR spectra (Schmitt et al., 1998), for instance using the characteristic absorption feature at 1.5 µm (Bibring et al. 2004). The depth of this feature is measured by the "BD1500" spectral criteria.

## 2.2 Spectral unmixing modeling

The shape of the 1.5 µm band of $H_2O$ ice is significantly controlled by temperature (Grundy and Schmitt, 1998) and grain size (Douté and Schmitt, 1998). Also, liquid water is spectrally distinct from water ice, with an absorption band at 1.5 µm shifted by 0.05 µm toward shorter wavelengths (Fig. 11). Other ubiquitous hydrated chemical species on Mars, such as sulfates, also have a band around 1.5 µm (Langevin et al. 2005, Bibring et al., 2006, Fishbaugh et al., 2007, Massé et al. 2010). The possible presence of water ice with different grain size, liquid water and gypsum are therefore considered while conducting the spectral analysis.

We applied a strategy of sparse linear unmixing using the Fully Constraint Least Square (FCLS) suitable to unmix spectral mixture in the case of an overcomplete reference database (Heinz and Chang, 2001). Each observed spectra $Y_k$ (k is the index of wavelength) is decomposed into a weighted sum of all reference spectra $S_{jk}$ (j is the index of the reference spectra, j=1 for $CO_2$, j=2 for $H_2O$ at 1 microns grain size, etc). The weight, or abundances, $A_j$ are assumed to be *summed to one* – because the sum of surface proportion is one - and *sparse* - with a distribution of zeros and only few non-zeros values.

$$Y_k=\Sigma A_j S_{jk} \qquad (1)$$

It means that if a spectrum is present in the database S but not present in the observed spectra Y, the method indicates a zero abundance. This property is especially suitable for our case because we do not expect all grain size of water ice, liquid water and gypsum (see later in the 3.5 section) appearing at the same time. Due to the sum-to-one constraint, we add a flat bright and a flat dark spectrum that can model at first order aerosol contribution and dust contamination at the surface respectively. Abundances can be interpreted as surface proportion inside the pixel, if no non-linear mixing occurs. In a more likely case, intimate non-linear mixing should occur, leading to a significant shift between absolute abundances and actually estimated abundances with FCLS. Nevertheless, we will interpret the abundances A as "spectral abundances". Such methodology is valuable to interpret global tendencies and for detection, but not for precise quantification.

### 2.3 Climate modeling

Observations of water and $CO_2$ ice at Richardson crater are compared with the predictions of the Global Circulation Model developed at the *Laboratoire de Météorologie Dynamique de Jussieu*. This model is able to predict the water cycle of Mars and has been validated against available datasets (Forget et al., 1999; Montmessin et al., 2004; Forget et al. 2008). The model computes the seasonal meteorology, notably the variations in water vapor and water ice precipitations, with a spatial sampling of 5x5° and a time step of half an hour. It is combined with a local energy balance code to predict the formation of ices on a given localized site with specific properties (Vincendon et al. 2010a, b).

## 3. Results

We first classify the observed features in section 3.1 using HIRISE images. Then, we analyze the spatial and temporal change in the properties of ice on the basis of their spectral properties as reveled by CRISM band ratio in section 3.2. In the section 3.3 we discuss how we assess the surface versus atmospheric origin of $H_2O$ spectral signatures. HiRISE and CRISM data are correlated in section 3.4. The results of the unmixing algorithm are discussed in section 3.5. Modeling prediction of ice condensation based on climatologic and thermophysic considerations can be found in section 3.6. Finally, we discuss the possible scenarios for the origin of the observed features in section 3.7.

**3.1. Classification of observed albedo features**

There is a great number of various seasonal low albedo features on Mars in the polar region, including different dark spots (Malin et al., 1998; Malin and Edgett, 2000; Zuber, 2003, Kieffer et al., 2006; Kossacki at al., 2006), spider-like structures (Kieffer, 2000; Ness and Orme, 2002; Piqueux et al., 2003; Christensen et al., 2005) Dalmatian spots (Kossacki and Kopystynski, 2004), and Dark Dune Spots (DDSs) (Horvath et al., 2001; Ganti et al., 2003; Pocs et al., 2003). Here we analyze only one group, called Dark Dune Spots, and among them only the larger ones, which different sub-units could be resolved on CRISM images. There are various subsections of these spots, analyzed recently (Kereszturi et al. 2009, 2010) based on HiRISE images. Here we use the following classification:

- Dark features on the seasonal frost cover, with two groups (Fig. 1): 1. large spots (larger than about 20-40 m) are present only between the ridges of the dunes on horizontal surfaces, and 2. smaller spots, which are present on the dune ridges and between them also, and on the slopes where flow-like features emanate from them (Horváth et al. 2009, Kereszturi et al. 2010).

- Here we analyzed only the large spots, large enough (i.e. > 60 m) to identify the following different sub-units of them from CRISM data:
    - *Dark core*: a roughly isometric and darkest area in the center, with albedo between 0.15-0.26 when the feature was evident before $L_S$ 250°. After this date the features diminished and the terrain becomes homogeneous.
    - *Gray ring*: a wide ring-like feature that surrounds the core, its albedo is intermediate between the dark core and the undisturbed frost cover, with albedo values between 0.25-0.30.
    - *Bright halo* (white collar): a region surrounding the spots brighter than the undisturbed average ice with albedo values between 0.37-0.24, somewhere elongated in downwind direction, possibly formed by refreezing process.
    - The *undisturbed ice* terrain surrounding far away the spots was also analyzed for control measurement.

These features show characteristic changes: first the dark core appears with dark wind-blown features emanating from it; later gray rings develop, then bright halos appear and finally the whole terrain defrosts (Fig. 2).

### 3.2. Distribution and properties of ices from spectral analysis

Three typical spectral CRISM observations of the 4 units (dark core, gray ring, bright halo, undisturbed ice) acquired during early to mid spring are analyzed in Figure 3. At that time near-IR $H_2O$ ice signatures are observed at every location, even in the darkest cores, while based on the optical images no substantial surface ice is present. Very small quantity of water ice (a few µm thick) could produce observable signatures in the spectra without visible effect in optical images, and thin water ice clouds may also contribute to the observed signature. Moreover, dust contamination within the ice (e.g., if ice as condensed on a dust nuclei) can significantly reduce the albedo of ice, making it barely detectable in the optical while clearly seen in the near-IR (Kieffer, 1990; Vincendon et al., 2007). $CO_2$ ice, detected on the basis of the 1.43 µm band depth, is also present everywhere at that time. It has been shown by Langevin et al. 2008 that aerosols scattering is responsible for a mixing of surface spectral properties when remotely observed, which could contribute to some of the signatures observed in this region characterized by high spatial variations of surface properties at the 10s to 100s meters scale. This effect, often neglected, can however have a major impact on observations as surfaces surrounding a km size observed target could contribute to several 10s of % of the observed signal (Langevin et al. 2008).

The spatial distribution of $H_2O$ and $CO_2$ ices is outlined in Figure 4. This figure contains small insets of HiRISE PSP_00307_1080 (1 column) and CRISM image FRT000052BC_07_IF163L (2-4 columns). Different rows are for different locations inside Richardson crater. The distribution of the water ice band depth (BD1500) in the second column shows that water ice is located in the grey outer ring area of spots. The $CO_2$ ice distribution is represented in the fourth column, showing that the darks spots are nearly $CO_2$ ice free, while the third column shows the areal distribution of the $H_2O/CO_2$ ice ration (inverted ICER1 channel) which highlights that. the outer gray ring is water ice rich. The ice ratio is a bit lower in the dark central core than in the ring, but here based on the optical images very small amount of ice could be present and the atmospheric dust scatter may have a strong impact in the spectra. While assessing changes in the composition of the ice mixtures from spectral analysis required the use of surface radiative transfer model (see e.g. Douté and Schmitt 1998; Skhuratov et al 1999; Hapke 2008 and reference herein), observing change in the ice band depth provides a first order idea of change occurring on the ground, keeping in mind that such changes can be due to several factors including grain size, photometry, and mixture type. $CO_2$ ice is nearly always present together with $H_2O$ ice, suggesting they could be mixed. But their ratio is changing, and could be correlated to the change in the appearance on visible images. $CO_2$ ice signatures are deeper in the bright halo and the undisturbed average icy surface, than in the darkest cores and the gray rings. The bright outer halo has the deepest $CO_2$ signatures compared to the average surface. The gray ring unit is relatively $H_2O$ rich and $CO_2$ poor, comparing their ratio to other unit types.

We highlight in Figure 5 the typical band depth change observed with time:

Changes of $H_2O$ ice based on BD1500 values:

- The different surface units show the same trend during $L_S$ 170° and 225°, except for the gray ring area.
- In the gray ring area the band depth increases from $L_S$ 180° until 220°, which is not observed for the other surface types, where nearly constant values are present.
- The water ice band depth is significantly lower for the dark core and bright halo in comparison to undisturbed terrain.
- After $L_S$ 225° there is a general decrease of the water ice band depth in all units which become spectrally similar.
- Water ice is confidently detected above the noise limit (~ 2%) until about $L_S$ 250°.

Changes of $CO_2$ ice based on BD1500 values:

- The depth of BD1430 increases from $L_S$ 170° until 225°, and then decreases.
- The bright halo feature shows stronger $CO_2$ ice signal than the other features, suggesting some re-condensation happened there after $L_S$ 210° (Titus et al., 2007).
- The $CO_2$ ice band depth is significantly lower for the dark core in comparison to undisturbed terrain.
- The $CO_2$ ice band depth is lower than the noise after about $L_S$ 250°.

Between $L_S$ 210° and $L_S$ 230°, the water ice band depth is higher relative to earlier and later periods in the gray ring unit. We discuss in the next section why this increase can be confidently attributed to surface bound ice.

### 3.3. Separation atmospheric and surface $H_2O$ ice signatures

An important but difficult issue is to separate water ice spectral signature of the surface from those of ice crystals in the atmosphere. There are attempts to separate the two sources by spectral calculation using the size dependent spectral properties of water ice in the 3 μm range (Langevin et al., 2007; Carrozo et al., 2009). However, no methods provide firm separation yet.

Another approach relies on the analysis of the spatial distribution of $H_2O$ ice signatures. Localized increase of the ice signatures associated with specific tens of meters scale morphological features is indicative of a surface origin of the ice signature. While conducting this kind of analysis, one should however carefully consider the role of potential photometric effect: the apparent spectral signature of a widespread ice cloud can also change as a function of the surface type over which it is observed. For example, ice cloud signatures can be enhanced when observed above a tilted surface in shadow, as such a surface will be mainly illuminated by light scattered by the atmosphere, which contain the ice cloud signature. In Figure 6, we show that the spectral signatures of $H_2O$ ice observed in the gray ring area can be confidently attributed to the surface, as it localized, repetitively observed for each spot, not associated with lighting conditions, and as no suspicious trends with albedo are observed (both the darker and brighter areas around the grey ring show lower water signatures).

### 3.4. CRISM - HiRISE correlation

We analyzed in detail the HIRISE image PSP_003175_1080 which is characterized by a high spatial resolution, a low noise, and a low solar elevation above the horizon (to make long shadows). It has been obtained at $L_S$ 210° (mid spring). Although it is not easy to interpret the illumination based topography of the terrain with changing frost cover in springtime, we think a small topographic step along the perimeter of the gray ring is detectable in the image: the surrounding brighter ice is elevated, while the area of the gray ring (as well as the darkest core) is at lower topographic level. We are now going to detail the analysis of dark shadowed and bright illuminated sections of the terrain that lead to this result.

To find out the topography from illumination, both the intrinsic albedo of the surface material and the exposure of the surface slope toward the direction of solar insolation should be taken into account, as both may contribute to the observed brightness of any point of the images. We identified five surface unit types based on relative brightness differences, labeled from 1 (the brightest) to 5 (the darkest) (see Figure 7). This analysis was realized on the above mentioned PSP_003175_1080 image with good spatial resolution. The following brightness categories were identified:

1. the brightest unit (DN values 600-530). This unit type covers only small part of the images and is located always right next to unit 2 (see below). It is located along the perimeter region of the spots. Members of this group are usually elongated roughly from the lower left to the upper right, in right angle to the direction of the solar illumination.
2. average brightness unit (DN values 580-460). This unit makes up most of the terrain surrounding the dark spots and probably consists of $CO_2$ ice.
3. moderately bright unit (DN values 440-300). It covers the outer gray ring of the DDS,
4. darker unit (DN values 460-240). Often elongated from the lower left to the upper right, also in right angle to the solar illumination. Besides the elongated ones this unit is also present on the left or top-left side of small, isolated and nearly isometric patches of unit 2.
5. darkest unit (DN values 130-40). It is present at the center of DDSs, where probably the barren, defrosted dark basaltic dune surface is visible.

We have derived the following conclusions based on the relative location of these brightness units:

- unit 1 and unit 3 frequently run parallel to each other on the right and left side of unit 2 respectively (Fig. 8. a1, a2, b1). Such appearance is expected if they are the solar facing (1) and shadowed (3) units respectively, produced by elevated and elongated $CO_2$ ice stripes,
- when units 1 and 2 form isolated, nearly isometric patches, unit 3 is also present at the left or top-left side of these brighter $CO_2$ “icebergs”, also suggesting unit 3 is composed of shadows,

- unit 5 is only present in the center of DDSs, which represent a feature independent of the $CO_2$ ice cover, and probably show barren dark dune surface without any kind of ice.
- unit 4 covers only the outer ring of the DDSs where unit 1 or 2 are present, and is not present in the immediate surrounding of the unit 5. It also strengthens the possibility that it is the shadow of elevated $CO_2$ ice features.

To summarize, as unit 4 is always on the left or top-left side of unit 2, and it is also often elongated parallel to unit 1, unit 4 probably represents the shadow of thick $CO_2$ ice (Fig. 8. a1, a2, b2). In this case the shadows could be used for the rough determination of thickness of $CO_2$ ice. For such measurement it is important to note that at some locations the elongated heights are composed of both the solid dune ripples below the $CO_2$ ice cover and the ice cover itself together (like Fig. 8. a1, a2). In these cases the height measurement would show not only the thickness of $CO_2$ ice but also the height of the dune ripple. But there are many examples where small isolated heights (such as units 1 and 2) and their shadows (unit 3) are present together, and no ripple-like feature is visible in their surroundings (Fig. 8. d1, d2, f1, f2). In these cases the heights are probably only the results of the unit 2 composed of $CO_2$ ice, and here the height estimation shows the thickness only of the $CO_2$ ice. It is also visible that bright units and darker shadows are present together close to the circular edge of dark spots, as one would expect if the $CO_2$ ice cover of the surrounding terrain begins here.

It is also possible that at some locations the surface albedo changes producing the difference in brightness, e.g. in the case of Fig. 8a1 where depressions on the left between dune ripples could be darker because of accumulated dark dust. But as the solar facing side of the bright unit is also present there, real topographic undulation should be present with solar facing bright slopes, and also with opposite facing shadowed slopes. And there are such cases (like Fig. 8. b2, c1, d1, d2, f1, f2) where it is very improbable that dust accumulated only on the shadowed sides of isolated bright patches. More probable explanation is that these dark features are shadows.

The shadow-like unit 3 always surrounds small isometric or larger elongated heights and always elongated toward the left or top-left. During the acquisition of the image the Sun was 24° above the horizon toward the lower right. The shadows' average length is about 20-30 cm, so the bright ice layer covering the whole terrain except the spots is roughly about 10 cm high.

To exclude the effect that earlier formed and permanent depressions are visible in the images (and not annually reoccurring features) we compared images acquired before the spot development with later images contain dark spots. On the earlier images PSP_002186_1080 and PSP_003175_1080 of the same area, acquired during continuous $CO_2$ frost coverage (at $L_S$ 166.20° and 210.6° respectively) no spot or any topographic depression is visible. The lack of such depression suggests that the gray rings' area was covered originally with the layer that was at least 10 cm thick at that time, which disappeared later and formed the depression; the original thickness could have been even larger, as $CO_2$ ice may have already started to sublimate at the time of the shadow measurement. Unfortunately the earlier image's resolution is substantially lower than the analyzed one, and all later images showing summertime defrosted surface were acquired under bad illuminating conditions, and they can not be used to analyze the topography in details. Figure 9. a) shows the earlier image of a terrain in Richardson-crater with complete $CO_2$ ice coverage, the b) shows the later image with several dark spots. The same boxed areas are magnified in c) and d) subsets. The continuous ice cover is visible in c), while $CO_2$ ice free, but still $H_2O$ ice covered areas can be seen in the d) subset. Arrows show dune ripples, which are present on the $CO_2$ ice covered and in the $CO_2$ ice free terrain too. The shadows by the CO2 ice cover are marked with arrow heads in subset d). It can be seen that even the topography of the ripples might affect the shadow length, the $CO_2$ ice layer definitely cast shadows. As it was shown earlier (see Figure 8.), there are such small, isolated, nearly isometric $CO_2$ heights, where no dune ripples below them affected the height of the $CO_2$ ice, and the thickness estimation based on shadows. The gray ring area is not barren dark opposite to the central core, suggesting some ice cover is still there. The gray layer is substantially thinner than the brighter $CO_2$ ice.

The location of the deepest $H_2O$ ice signatures that could be confidently attributed to the surface (see section 3.3) based on CRISM data coincides with the gray ring features seen in the HIRISE data (Fig. 10). In this area the small-scale dune ripples are also visible. This observation is compatible with the idea that condensed $H_2O$ covers the surface there in a thin layer, which allows the topographic undulation of the surface to remain visible. The grey ring area also contains weaker $CO_2$ ice signatures. It has been shown by Langevin et al. 2008 that aerosols scattering are responsible for a mixing of surface spectral properties when remotely observed. In particular, an ice free area of limited extent (< km) surround by a widespread ice rich area will contain ice signatures when seen from orbit as a result of the mixing of spectral properties by aerosols scattering. Surrounding icy areas will typically represent 10 to 30% of the observed signal (Langevin et al. 2008). As a consequence, the grey ring area could be free of $CO_2$ ice as it is of small extent, characterized by the weakest $CO_2$ signatures of the area (see Figure 4 and Figure 6) and surrounded by terrains with strong $CO_2$ ice signatures. Moreover, in Figure 11 we show HiRISE images of the dark spots showing bright patch consistent with ice located within the dark spots. These cm-dm size patches are below the CRISM resolution. As a consequence, spatial mixing could also contribute to the $CO_2$ ice spectral signature seen by CRISM in the dark spots and grey ring, which could be mainly $CO_2$ ice free in mid spring.

### 3.5. Spectral unmixing results

Using the FCLS linear unmixing method, we study the properties of this ice layer and we notably test the possible presence of liquid water in the four identified units. We consider 6 endmembers spectra (water ice with 3 grain size, $CO_2$ ice, liquid water and gypsum) which are representative of the expected potential surface components and which have a significant impact on the 1.5 µm spectral regions (see Figure 12 and section 2.1). $CO_2$ and $H_2O$ ices have been modeled by a radiative transfer simulation (Douté and Schmitt, 1998) using optical constant recorded in the laboratory (Schmitt et al., 1998). Both gypsum and liquid water have been acquired at Labotaroire de Planétologie de Grenoble (Schmitt and Pommerol, unpublished). These spectra will be soon available online at http://ghosst.obs.ujf-grenoble.fr. Two spectrally blank spectra are also considered to account for dust and aerosols. Also an atmosphere transmission spectrum has been added to take into account, small residue of the atmosphere removal.

We estimate the abundance of each reference spectra (see Fig. 12) for the CRISM images: 43A2 ($L_S$=181.7°), 52BC ($L_S$=213.7°), 56CO ($L_S$=221.0°), 5C94 ($L_S$=241.7°), 5E38 ($L_S$=245.5°), 5FF6 ($L_S$=248.7°) and 6516 ($L_S$=262.6°). We used the range 1.04-1.38 µm and 1.51-2.13 µm to remove the narrow $CO_2$ feature that could be affected by the spectral smile. We also weight by a factor of 5 the spectral range 1.50-1.53 µm, very sensitive to water ice, liquid water and gypsum. The root-mean-square (rms) is then computed without weight. All results for the model are shown in the Appendix A.

The general trend of the total $H_2O$ ice spectral abundance (sum of $H_2O$ with grain size of 1 and 100 μm) is compatible with the band ratio (Fig. 5 top). Water ice at 1 cm grain size is never detected. The relative fraction of 1 μm grain size is plotted in Figure 13. The dark core has the smallest mean grain size of the four units. The higher grain sizes are identified in the undisturbed terrain. Figure 13 suggests a general trend with increasing grain size until $L_S$= 220° and then decreasing until $L_S$=240°. These facts indicate that: (i) water ice is present with relative low grain size, consistent with seasonal ice; (ii) the grain size is evolving with time, which can be due to several mechanisms including resurfacing by sublimation, recondensation during the night, sublimation or deposition by geyser activity, grain metamorphism.

The results show that the spectral abundance of $CO_2$ ice is decreasing from 10% to 3% in agreement with band ratio BD1430 (Fig. 5 bottom). Also, the fit of $CO_2$ ice bands is compatible with large grain size (10 cm) and possibly a translucent slab ice. The last observation 6516 shows an estimated abundance of 0.06%, probably due to atmospheric residue.

We have also included liquid water and gypsum – a mineral with a 1.5 microns band resembling that of liquid water - in our modeling procedure. In Figure 14, we show one example of the model results for the dark spot of image 56CO for three linear unmixing scenarios: (1) without liquid nor gypsum (2) with liquid water only (3) with liquid water and gypsum. The rms is decreasing from model 1 to model 3 due to a slightly better fit of the 1.5 microns band (see Figure 14). This tends to indicate that a spectral component with a water band shifted toward shorter wavelengths, such as gypsum or liquid water, could be present. Sulfates and/or liquid water have been detected in six over the seven images considered (i.e.: for all observation except the last one at $L_S$=262.6°) in the “dark center” region, while it is almost never detected in the average terrain or in the bright halo. However, retrieved abundances are low (< 3%, see Appendix A), and differences in modeled spectra are not significant given the first order method employed here (Figure 14). Although the retrieval of a few % of species with a 1.45 microns band consistent with liquid water in the dark spots only is intriguing, no robust conclusions can be reached. Moreover, none of the ratio spectra show evidence for liquid water or gypsum (see Fig. 15). To assess more precisely the possible presence of liquid water and/or gypsum in CRISM spectra, a non-linear spectral inversion using radiative transfer model should be performed in the future.

### 3.6. Climate modeling results

The surface stability and presence of ices are analyzed at the latitude of Richardson crater as the season progresses from winter to summer using a climatic model (see section 2 for details). Three modeling hypotheses concerning local surface properties are shown in Fig. 16: (A) diamonds for standard parameters (flat surface, moderately bright ices), (B) stars for condensation on a 15° pole facing slope (to estimate the effect of local favorable conditions), (C) triangles for "dark" ice (albedo of 0.25 for both $H_2O$ and $CO_2$ ice, a rough attempt to assess the impact of dust contamination of ice). These modeling predictions are compared to the presence or absence of ices at Richardson crater as a function of season as constrained from CRISM data, which show a general good agreement between data and model.

Water ice accumulation starts before $CO_2$ ice at the beginning of fall, mainly due to snow precipitation of small grain size particles (more than 99% of the total water ice accumulation) that have condensed in the lower atmosphere according to the simulation (see also Forget et al., 2006). A thin layer of a few microns to a few 10s of microns thick $H_2O$ ice below the CO2 ice is expected according to model results, as noted by Schmitt et al. 2009. Then both ices condense simultaneously ($H_2O$ inclusion within the $CO_2$ ice) 20 to 30° of $L_S$ later. As a result we have a layered structure: only $H_2O$ ice below, and $CO_2$ plus $H_2O$ ice above. The ratio of $H_2O$ ice in the upper layer depends on several factors, above all meteorological conditions and global atmospheric circulation, as during the $CO_2$ condensation in southern winter the average atmospheric vapor content reaches maximum supplied by the exposed northern permanent water ice cap. When cold $CO_2$ ice is present, it traps available water vapor. Ice disappears prior to the summer solstice even for favorable conditions. According to the model, water ice remains 5° to 10° of $L_S$ later compared to $CO_2$. The contamination of ice by dust significantly decreases the stability of ice in spring, which sublimation is shifted by about 30° of $L_S$. According to the model, $CO_2$ ice disappears at $L_S$ 220° for a moderately dark ice with a 0.25 albedo (Figure 16). This is consistent with the morphological and spectral analyses conducted above, showing that at a similar period ($L_S$ 210°, see section 3.4) the grey ring is composed of an $H_2O$ ice rich thin layer while surrounding areas are still covered by a thicker layer of $CO_2$ ice.

### 3.7. Formation mechanisms of $H_2O$ in the gray outer area

Based on the observations and modeling predictions presented in the previous sections, we propose the following scenario for the formation of the observed surface $H_2O$ ice in the dark dune spots of Richardson crater:

1. During early fall, $H_2O$ deposits start to form at the surface. They are mainly composed of precipitations that have settled from the lower atmosphere. Several tens of µm of water ice accumulate at the surface (see section 3.6).
2. During mid-fall, concomitant condensation of $H_2O$ and $CO_2$ begin and form a layer of $CO_2$ ice contaminated by $H_2O$ ice. Several tens of cm of frost, mainly composed of $CO_2$ ice, form in the whole area during that period according the earliest CRISM observations, the HiRISE observations of a 10 cm depression in mid-spring (section 3.2) and the modeling predictions (see section 3.6).
3. After the spring equinox, dark core are observed at Richardson crater. They probably result from the formation of $CO_2$ jets below the translucent ice (Kieffer, H.H., 2000), which carry along dune grains that settle about the jet. This scenario is in agreement with the results from the linear unmixing method: (i) detection of large $CO_2$ ice grain size; (ii) higher abundances of dark dust for the dark core and grey ring units (see section 3.5). Around the dark spot a "grey ring" is observed, which spectral composition during mid-spring is consistent with an enhanced amount of water ice (see section 3.3) and probably no $CO_2$ ice. Two possible scenarios can then explain or contribute to the observed properties of this grey ring:
    a. The presence of dark dust material fallen from the jet increases surface absorption and is responsible for ice sublimation. Ice sublimates faster at the center of the spot, which explain the depression observed in HiRISE images and the weak $CO_2$ ice bands observed in CRISM data for which spatial mixing of ice-free and ice-rich subpixel deposits, as well as aerosols scattering, is a probable explanation (see section 3.4). In the outer grey area, albedo is higher and causes higher stability of water ice, as indicated by the model (see Fig.

15). The relative smaller water ice grain size could be due to the reappearance of the small grain size layer of snow precipitated in early fall.

   b. The water ice, that is present between the sand substrate and the translucent $CO_2$ ice layer, is blown up with sand grains by geyser activity (Kieffer 2000, Piqueux et al. 2003). Water ice grain falls on the top of the surface surrounding the central dark core, contributing to the formation of the water ice rich grey ring. The relative smaller water ice grain size could be due to the geyser activity.

4. Although no robust evidence for liquid water has been found in the spectra, the data are not inconsistent with the possible presence of small amount of liquid water in the dark core unit and, with an even lower probability, in the grey ring unit (see section 3.5).

## 4. Conclusions

We have performed a study of the Dark Dune Spots observed in Richardson crater using HiRISE and CRISM data. Our goal was to identify and analyze the presence of water ice within these features. It has been possible to analyze the largest spots using the 20 m spatial resolution of CRISM, which provided clues about the presence of water ice within these features. In particular, Dark Dune Spots can give rise to flow-like feature and the role of liquid water in this process is the subject of ongoing research.

Based on the analysis of HiRISE images, we identified three characteristic sections of Dark Dune Spots observed in Richardson crater during spring: dark core, gray ring and bright halo. We used time series of CRISM images between $L_S$ 175.5° and $L_S$ 340.5° to analyze temporal changes of $H_2O$ and $CO_2$ related band depth values. At the area of gray rings on HiRISE images the bright icy layer that surrounds the spots is missing. Based on shadow length measurements an about 10 cm deep horizontal depression is present there, measured relative to the top of surrounding bright ice cover. These depressions were not present in earlier images, suggesting it forms due to the localized sublimation of the ice layer.

We identified $H_2O$ and $CO_2$ ice at most locations during spring, but with significant spatial and seasonal variations in band depth and linear unmixing results. Spatial subpixel mixing, water ice cloud contribution, and aerosols scattering responsible for an apparent mixing of surface properties must be carefully considered while analyzing the apparent spectral properties in CRISM data. Enhanced water ice signature is observed during mid-spring at the gray ring area from BD1500 values. Based on the size of these ring features (20-30 m) and on the repeatability of these features from one spot to the next, the surface origin of this water ice has been confidently determined.

By analyzing CRISM images, HiRISE data and the result of a climate model, it has been possible to build a scenario for the formation and evolution of the ice layers in and around the dark dune spots of Richardson crater. During fall thin, tens of μm thick $H_2O$ ice rich layer forms at the surface. 30° of $L_S$ later a layer of $CO_2$ ice with inclusions of $H_2O$ ice forms on top of it, and reach a thickness of a few tens of cm. During late winter / early spring the dark spots form, probably by geysers activity (Kieffer, 2000) in agreement with the large $CO_2$ grain size detected by the linear unmixing. A bright halo-like feature surrounds the spot and show stronger $CO_2$ signature than the undisturbed average ice-covered surface, suggesting $CO_2$ was refrozen there. The dark spots and the grey rings are located in a 10 cm thick depression compared to the surrounding bright, ice-rich layer. The dark center spot is ice-free at some places (but contain ice signatures in CRISM images due to spatial mixing and aerosols scattering). At $L_S$=210-230°, the gray ring area is covered by an enhanced thin layer of $H_2O$ ice. The origin of $H_2O$ may be from two sources: (i) condensation on the surface in autumn (before the formation of $CO_2$ layer) or during winter time ($CO_2$ layer acting as a clod trap) and exhumed by the differential sublimation of $CO_2$ and $H_2O$ ice (ii) $H_2O$ attached to the sand grains that are blown up by geyser activity and that fall at the surface around the spot.

We observed $H_2O$ ice left behind the disappearance and/or reduction of $CO_2$ ice thickness in the gray ring area of Dark Dune Spots. Resembling situation was observed as a shrinking $H_2O$ ring during the recession of the northern seasonal cap (Apperé et al., 2008, Wagstaff et al. 2008), and in some patches at the southern hemisphere too (Brown et al. 2010), but at substantially larger spatial scale.

The presence of water ice in physical contact with insolated dark dune material could lead to the formation of liquid water. Such environments are of interest while assessing the potential for habitability of present day Mars (Szathmary et al. 2007). In the gray ring area the water ice surrounds darker surface, where liquid interfacial water layer or brine (Möhlmann 2004, 2009, 2010) may form. We found no firm evidence for the presence of liquid water in near-IR spectra, although linear unmixing results show that the data are not inconsistent with a possible slight contribution (a few %) of liquid water in the dark core unit.

The above-mentioned observations may have important impact on the explanation of other circumpolar features, showing water ice may be present in $CO_2$ ice free regions at the southern hemisphere of Mars. We observed the presence of water ice in physical contact with relative dark dune grains in 100-200 meter sized spots. In their surrounding of these spots in Richardson crater flow-like features are present emanate from resembling but smaller spots, which are too small to be resolved by CRISM. If resembling process happen there, such locations could be also interesting for the possible ephemeral presence of water ice, as well as other seasonal features should be analyzed in detail to search for water ice there.

## 5. Acknowledgment

The authors would like to than Bernard Schmitt and Francois Forget for their help. This work was supported by the ESA ECS-project no. 98076. This work was also supported by the Centre National d'Etudes Spatiales (CNES). It is based on observations with CRISM embarked on MRO. We also acknowledge partial support from the Programme National de Planétologie (CNRS/INSU).

## 6. Abbreviations

CRISM - Compact Reconnaissance Imaging Spectrometer for Mars, DDS - Dark Dune Spot, FCLS - Fully Constraint Least Square, HiRISE - High Resolution Imaging Science Experiment, $L_S$ - solar longitude, rms - root-mean-square, LPG – Laboratoire de Planétologie de Grenoble, FCLS – Fully Constrained Least Square

Figure captions

Fig. 1. The analyzed unit types of the spots in Richardson crater on the HiRISE image # PSP_003175_1080. Dark cores and gray outer rings are visible in the left subset, while elongated and probably wind blown bright halos are present in the right subset.

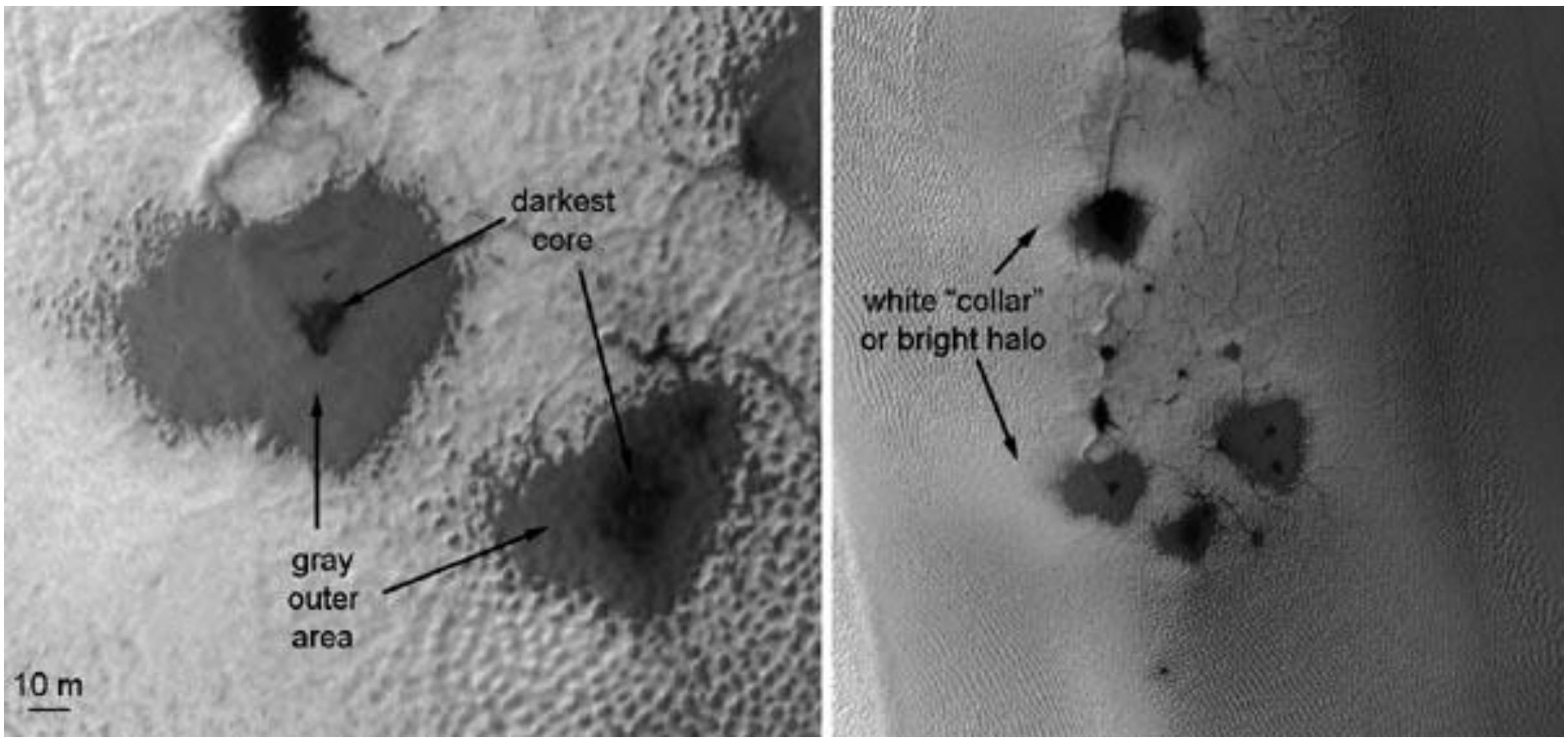

Fig. 2. Sequence of HiRISE images showing the temporal change of the same terrain with Dark Dune Spots in Richardson crater. At the earliest phase (a, b) only dark cores are present and wind blown dark streaks emanate from them toward the left, later the continuous gray ring forms around the dark cores (c, d), even later dark cores are expanding and gray rings contracting (d,e), finally only albedo differences are visible (f) but no ice is present based on CRISM data (image number and solar longitude values are: a) PSP_002186_1080 ($L_S$=166.2), b) PSP_002397_1080 ($L_S$=172.5), c) PSP_002885_1080 (Ls=197.0), d) PSP_003597_1080 ($L_S$=230.9), e) PSP_003742_1080 ($L_S$=238.1), f) PSP_003953_1080 ($L_S$=248.1)).

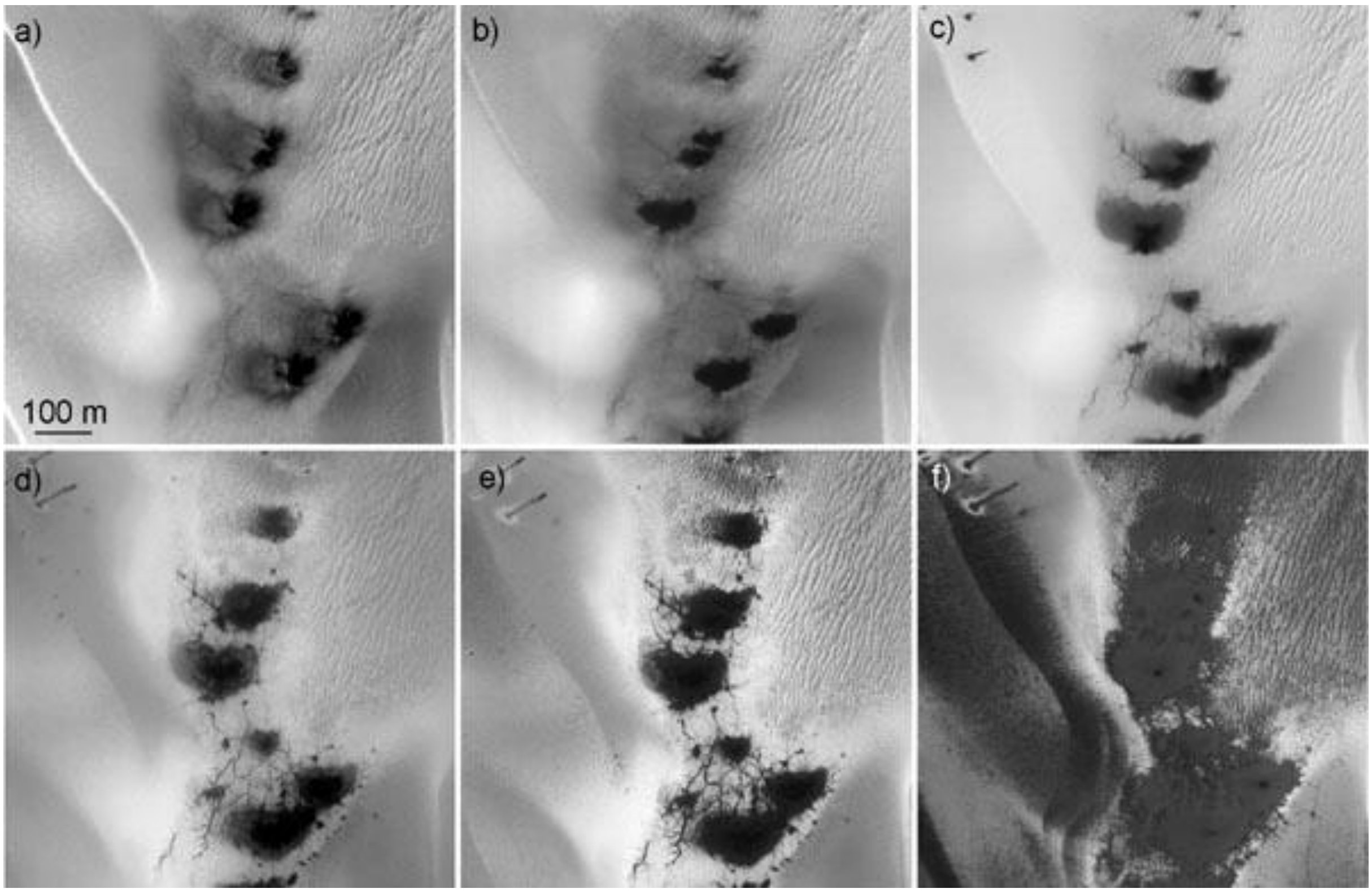

Fig. 3. Spectral shapes for three different periods (on images: 43a2, 5C94, 5FF6 at $L_S$=181.7, 242.7, 248.7), marked with black, gray and light gray color) for the analyzed surface units (dark centre, gray ring, bright halo, undisturbed ice), normalized to the continuum, between 1.0 and 2.6 micrometer. Carbon-dioxide and water ice signatures are present with decreasing band depth as the season progresses.

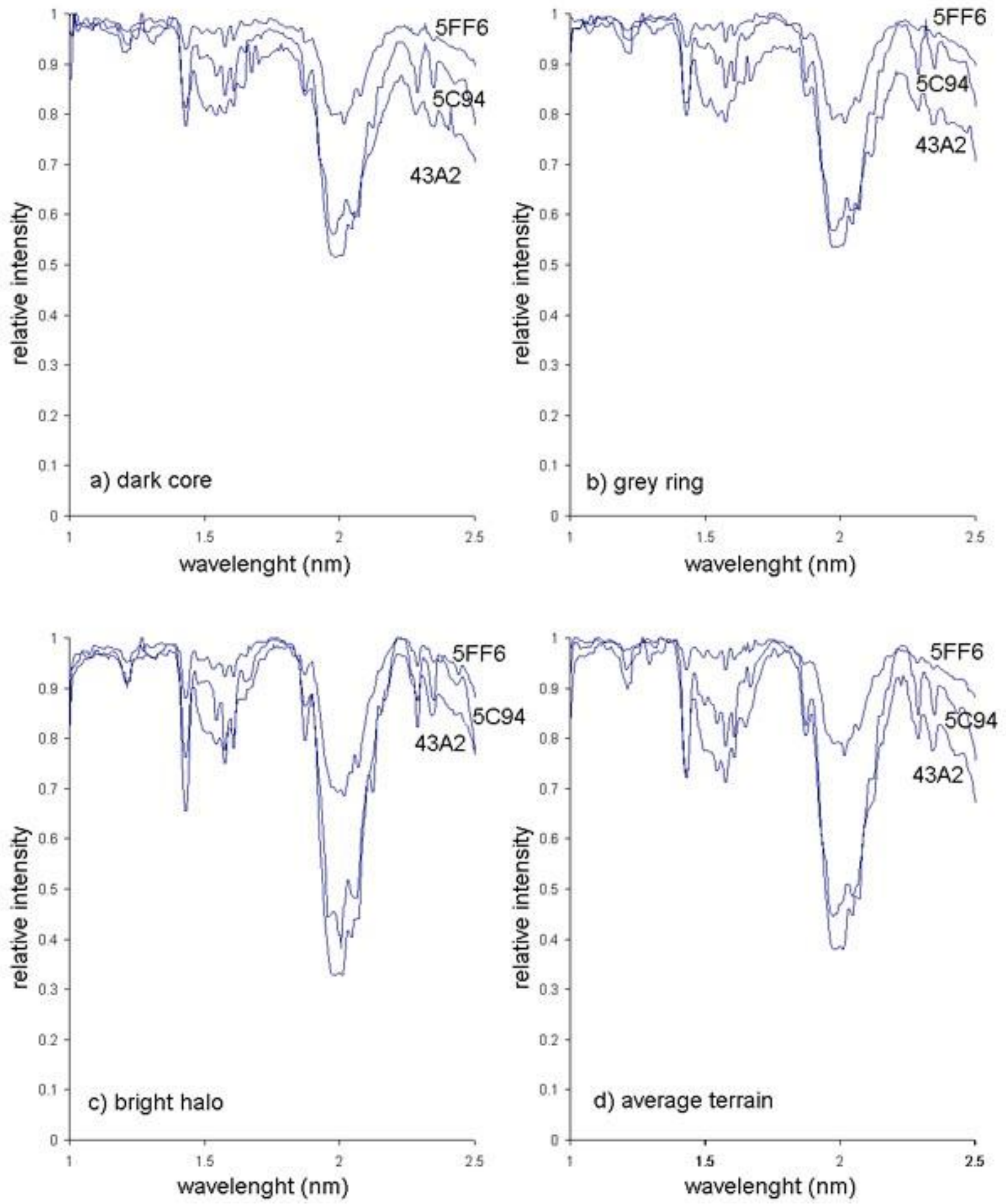

Fig. 4. Comparison of the distribution of water and $CO_2$ ice. The different rows show different locations in Richardson crater, and different columns are for: optical images (PSP_00307_1080 Column #1), water ice band depth image (BD1500 Column #2), image representing the ratio of $H_2O/CO_2$ ice (inverted ICER1 band Column #3), and $CO_2$ ice band depth images (BD1430 Column #4). The bright water ice rings in column #2 suggest surface $H_2O$ ice is present in the gray ring unit.

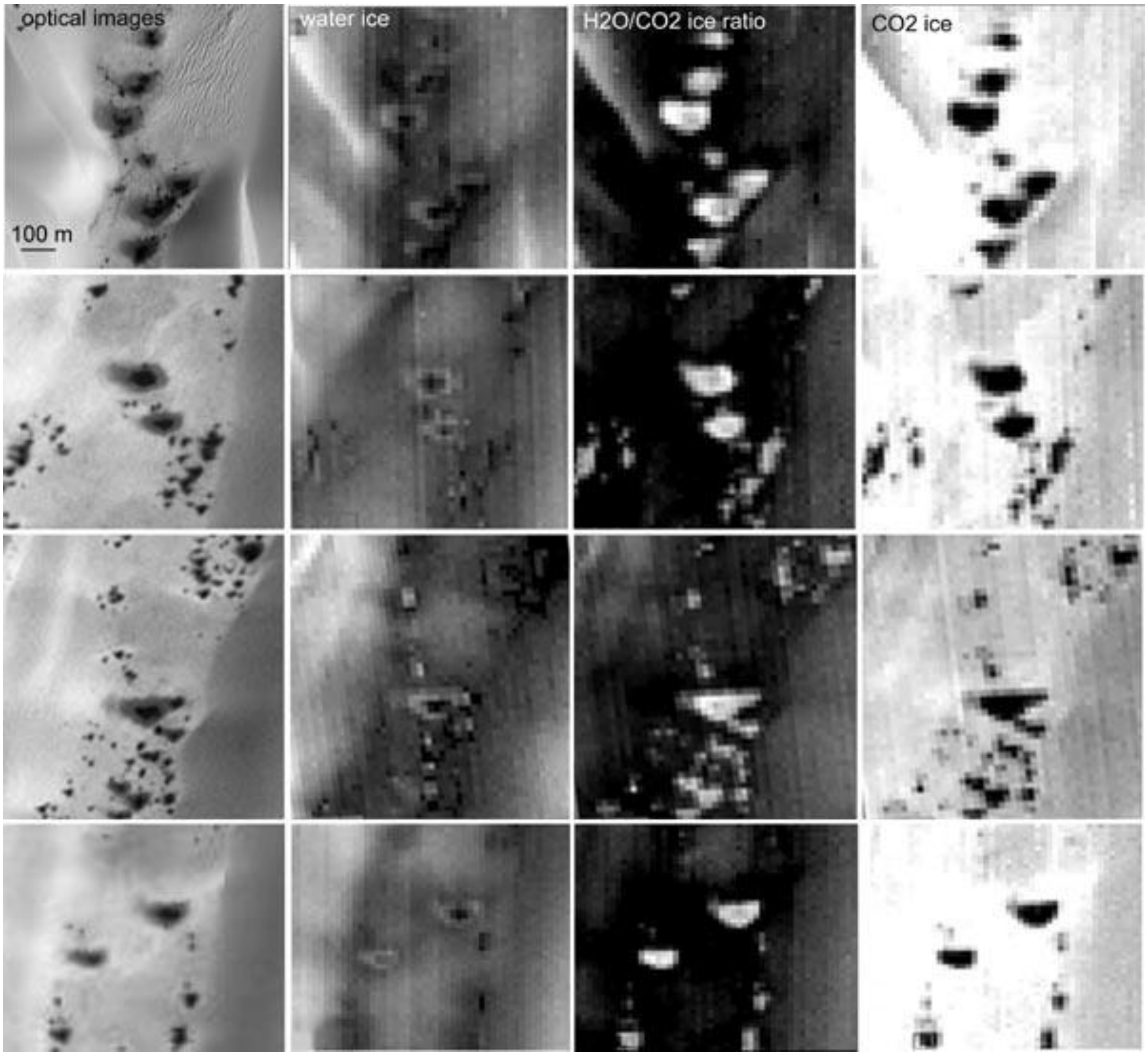

Fig. 5. Temporal changes of $H_2O$ and $CO_2$ ice related band depth values (on left) and spectral abundances estimated with the linear unmixing method FCLS (on right, see fig. 12 for more detailed explanation) for different units (core: blue, ring: pink, halo: yellow, undisturbed ice: blue). The top right panel shows the sum of the abundances of water ice at different grain size (only grain size of 1 and 100 microns are detected). The bottom right panel shows the abundance of $CO_2$ ice (only at 10 cm grain size). Top:  $H_2O$ ice. A general decrease can be seen, except for the gray ring, where an increase is present about Ls 215°. Bottom: $CO_2$ ice. The amount of $CO_2$ ice increases during early spring, with a stronger increase seen during the formation of the bright halo feature (yellow).

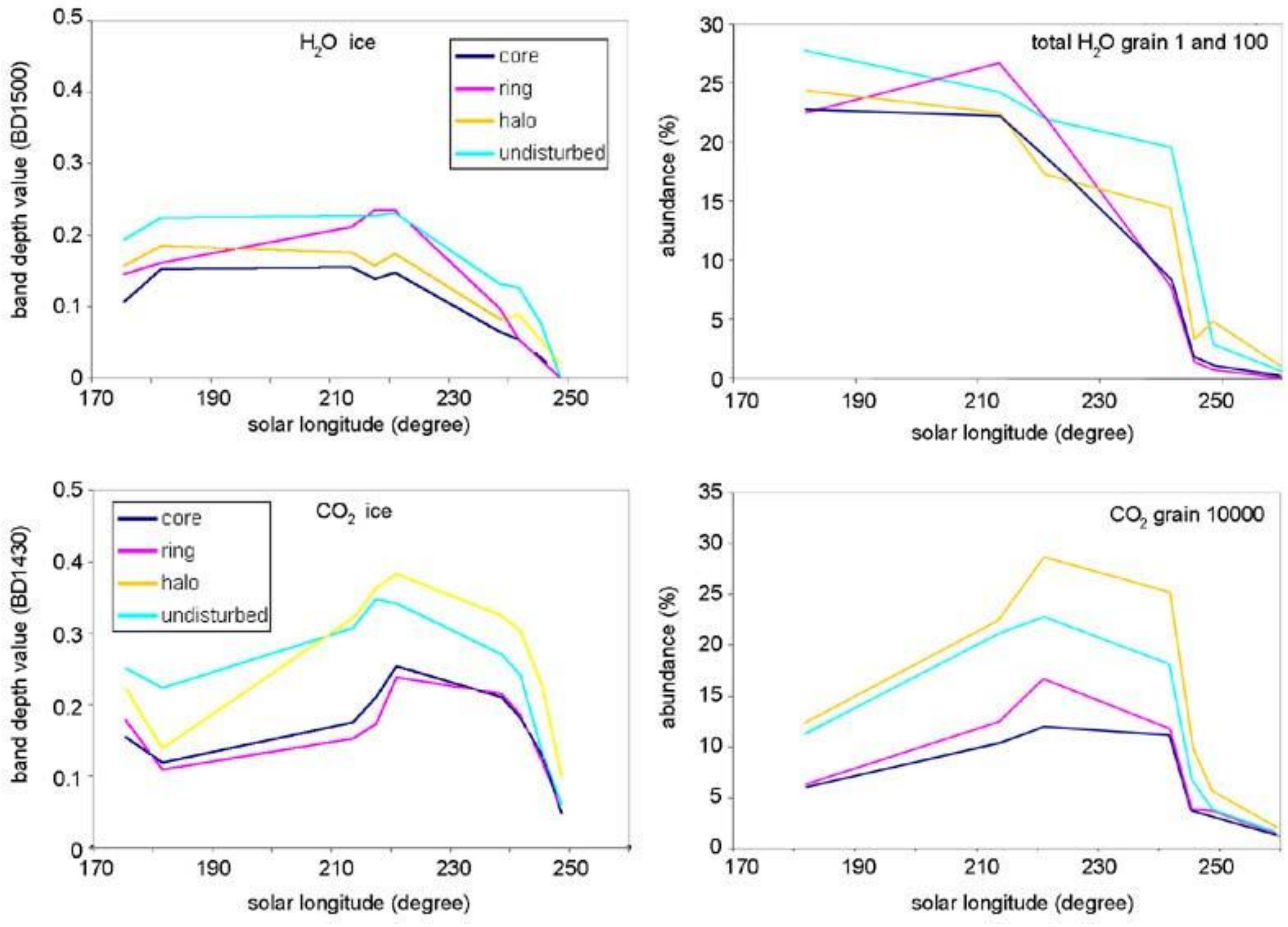

Fig. 6. Spatial changes on images (left) and graphs (right) along a spot for different parameters: a) albedo (HiRISE), b) water ice (BD1500), c) ratio of $CO_2/H_2O$ ice (ICER1) and d) $CO_2$ ice (BD1430). The two gray columns on the right mark the location of the gray ring (R) area along the profile, showing a decreased $CO_2$ ice signature and an increased $H_2O$ ice signature at this location. The magnified image of the analyzed spot is visible at the top right. The subsets show that the lowest albedo is found in the core (C), which is $H_2O$ ice poor and shows some $CO_2$ ice signature (from atmospheric dust scattering of outer surfaces and/or from the surface). The outer ring (R) corresponds to a local minima in albedo with an elevated $H_2O$ ice signature a low $CO_2$ ice signature. The high $CO_2$ signature of the bright halo (H) and the average frost covered terrain (F) is also marked, while slope effect (SE) is indicate the elevated value by a solar facing large slope.

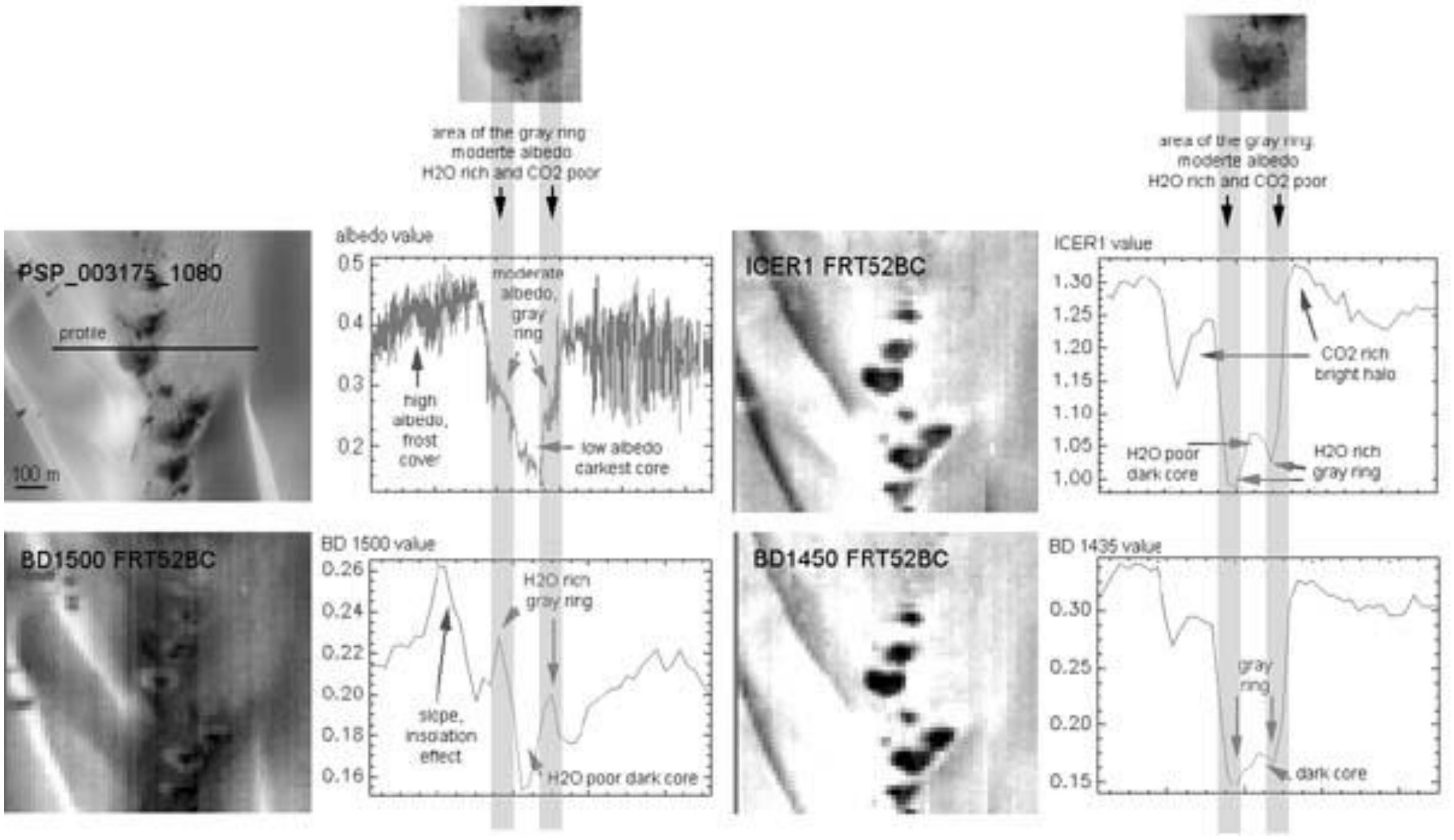

Fig. 7. Estimation of the topographic heights from shadow measurements (HiRISE image PSP_003175_1080). Five units are distinguished on the basis of their brightness (from 1 to 5). The context image a) shows a Dark Dune Spots with the locations of the magnified b) and c) subsets. In these subsets numbers arrows point toward examples for 4 of the 5 different units, while the darkest unit # 5 can be seen in the main image (a).

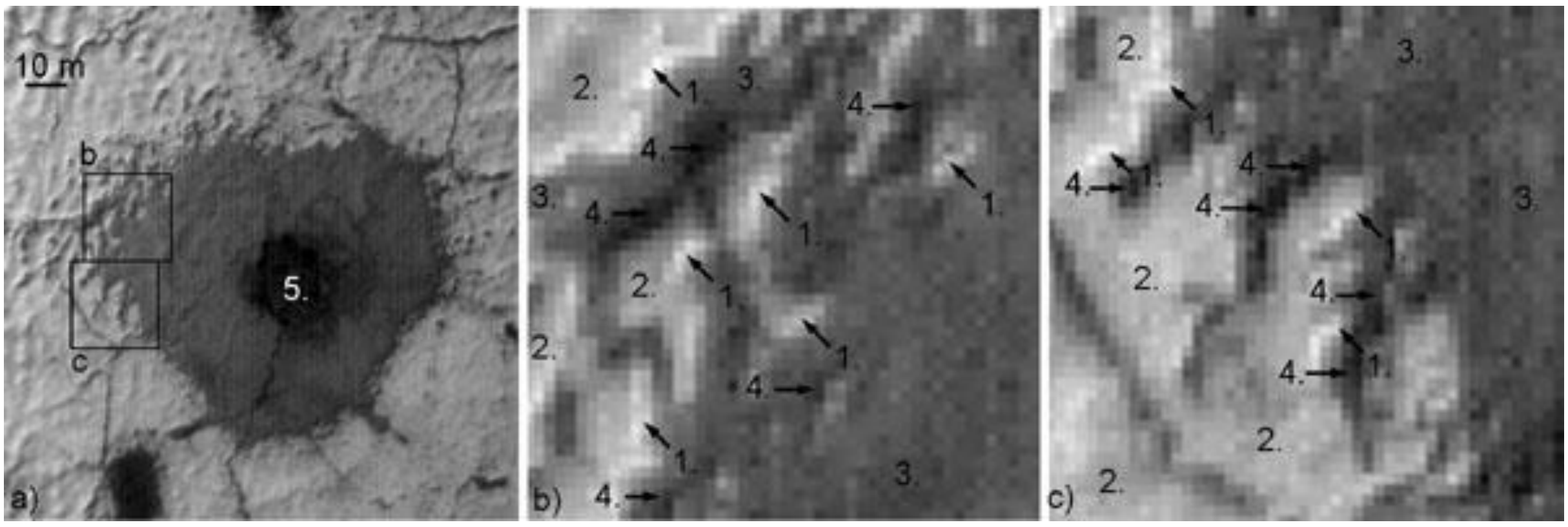

Fig. 8. Examples for the disappearing bright layer forming the depressed gray ring on the HiRISE image no. PSP_003175_1080. Six large spots are shown (largest images labeled a, b, c, d, e, f). Ten insets (10x10 m) from these images are extracted (a1, a2, b1, b2, c1, d1, d2, e1, f1, f2). Outlines of these 10 insets have been produced to highlight the main features and the shadows. In the small inset images the bright ice layer casts shadows, somewhere shadow are present behind iceberg like heights. The Sun illuminates from the lower right, shadows are elongated to the upper left. The outlines show the solar facing and brighter slopes with white, the opposite slopes in shadow with dark, and the nearly horizontal plains (between or on the top of the heights composed of $CO_2$ ice) with gray color.

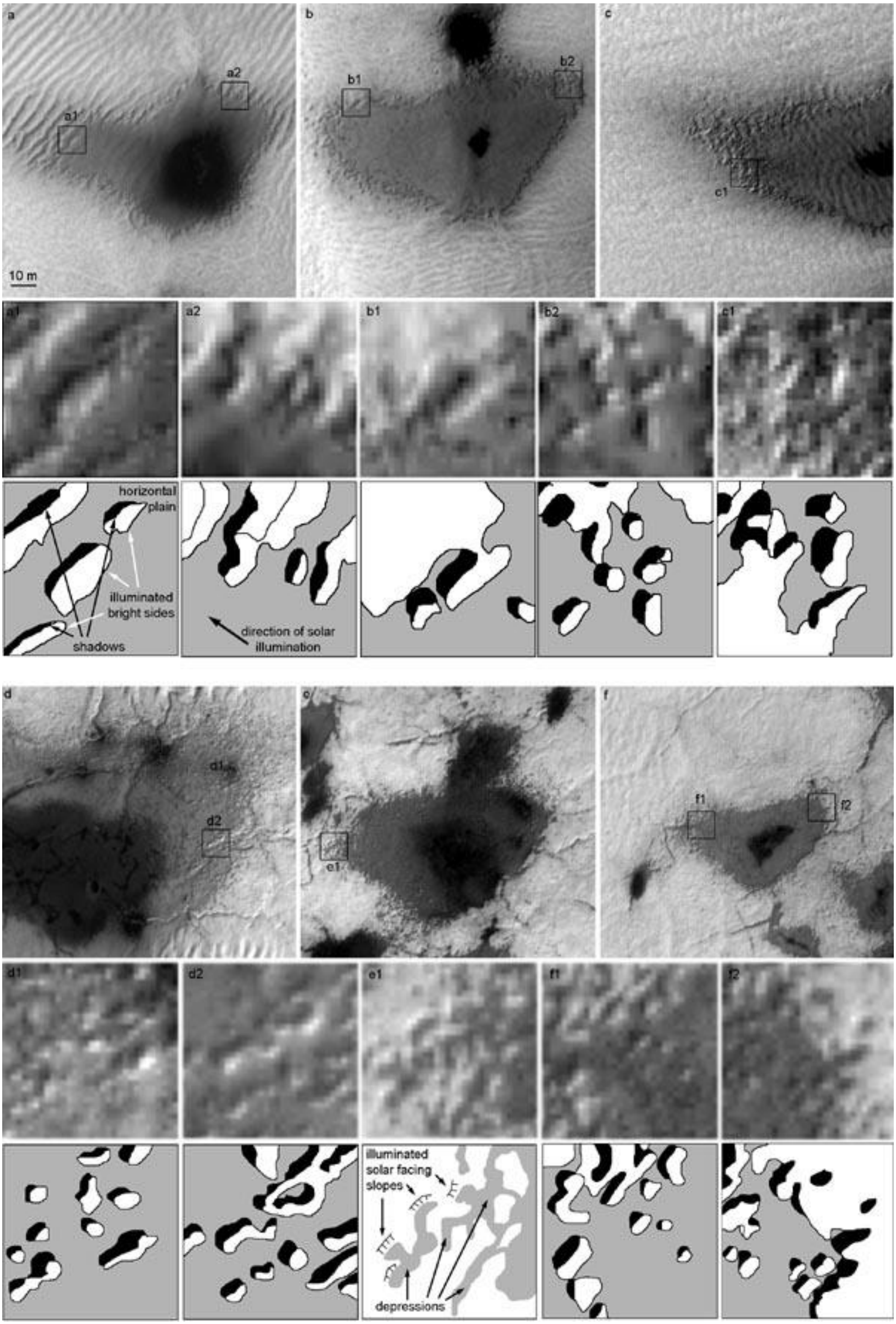

Fig. 9. An example area before and after the development of a spot in Richardson crater. The a) and b) image PSP_002252_1080 and PSP_003175_1080 were acquired at Ls=169.0, 210.6 respectively. The same small boxed areas are extracted below in c) and d) subsets. It can be seen that the depression of the gray ring (b, d) was not present earlier (a, c), in the previously $CO_2$ ice covered area. Small dune ripples are present in the $CO_2$ ice covered and $CO_2$ ice free area also. In subset d) the surface topography may contribute in producing the shadows, but the image suggests $CO_2$ ice contributed alone substantially in making the shadows, , as it also can be seen from isolated $CO_2$ ice heights in Fig. 8.

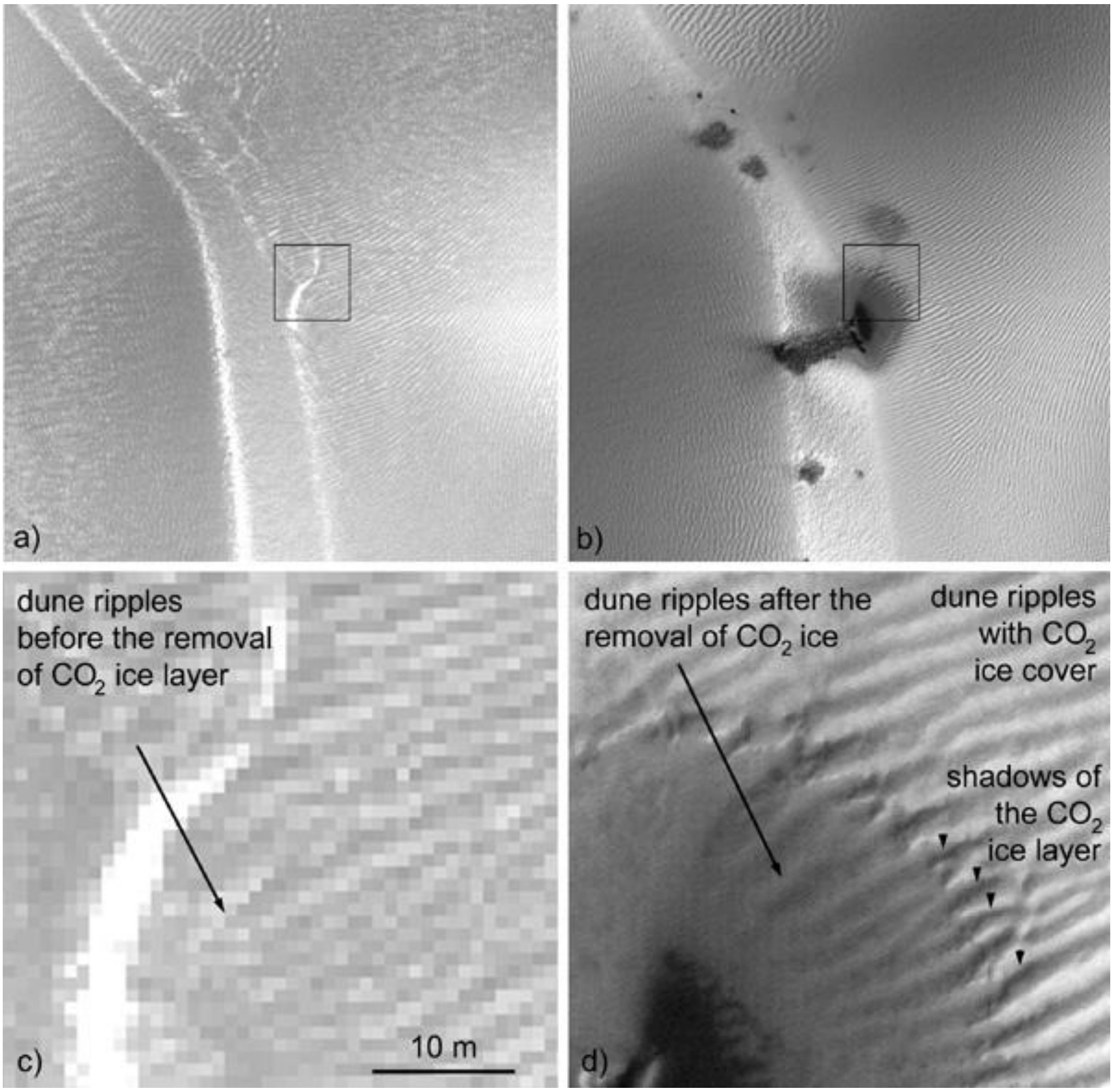

Fig. 10. Three examples for the spatial correlation between HiRISE image insets (left, from PSP_003175_1080 HiRISE image) and BD1500 values (right) from the CRISM image insets (FRT000052BC_07_IF163L). Despite the difference in spatial resolution we can see that the location of the strong BD1500 values (bright pixels on the right) coincide with the gray ring area (on the left).

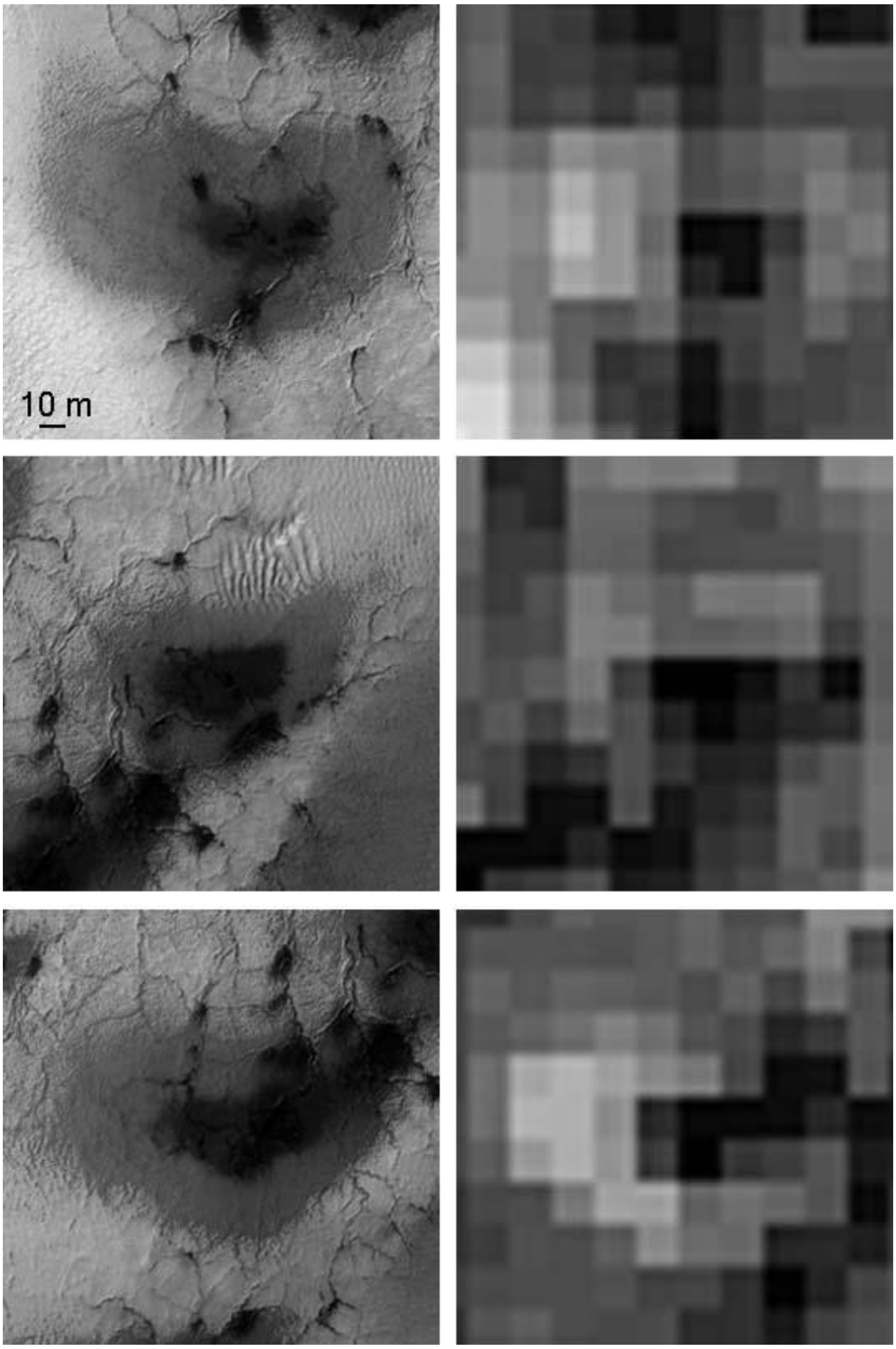

Fig. 11. Examples for small bright patches inside the spots (arrows) on HiRISE image PSP_003386_1080. These suggest, small $CO_2$ ice patches may still be present even at the darkest cores, which are below the spatial resolution of CRISM data, but may effect the spectra.

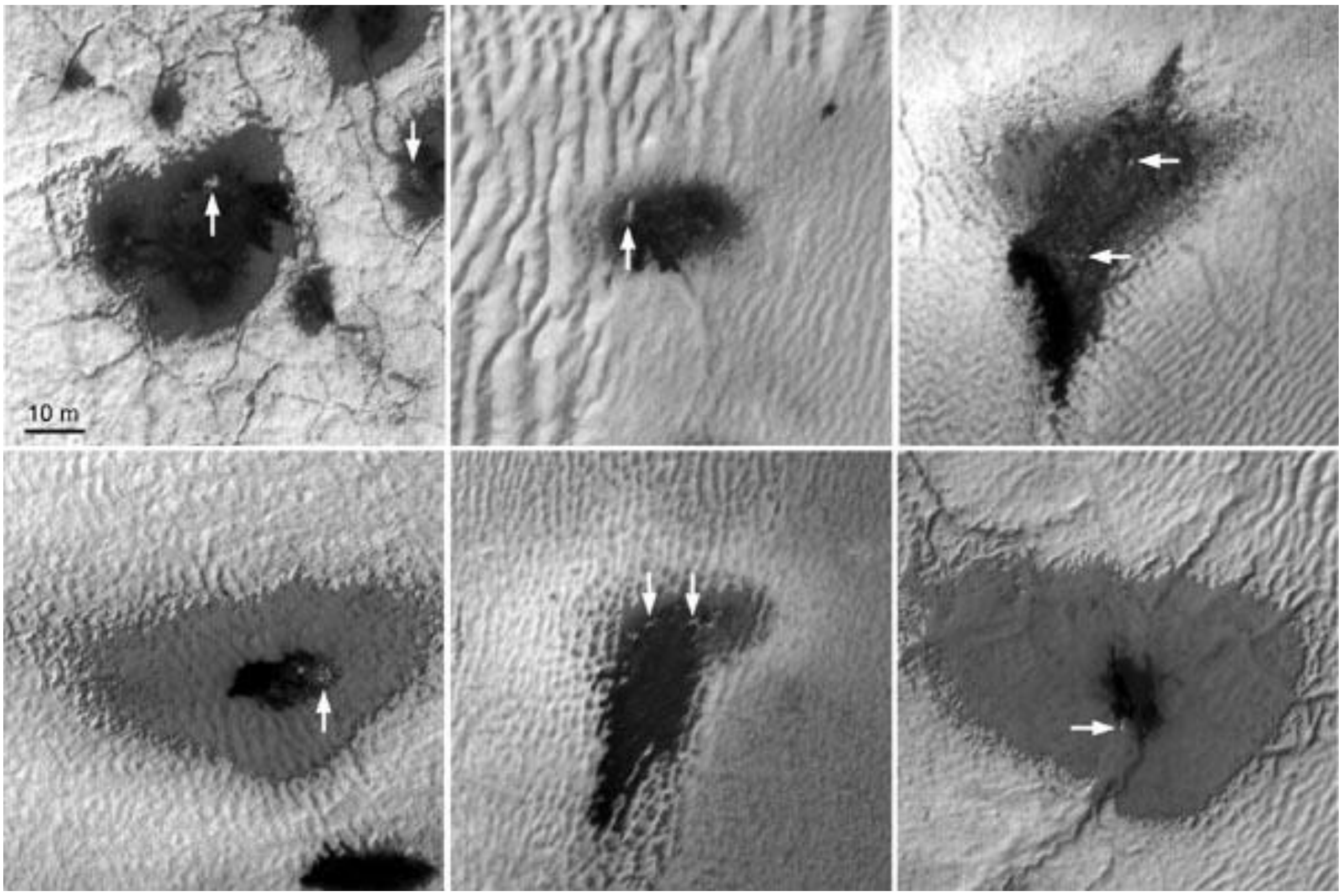

Fig. 12. Reference spectra used by FCLS to estimate the relative abundances for observed CRISM spectra. $CO_2$ ice at 10 cm grain size, $H_2O$ ice at 1, 100 and 1000 microns have been modeled by a radiative transfer code (Douté and Schmitt, 1998) using optical constant recorded in the laboratory (Schmitt et al., 1998). Both gypsum and liquid water have been acquired at LPG (Schmitt and Pommerol, unpublished). These spectra will be soon available online at http://ghosst.obs.ujf-grenoble.fr.

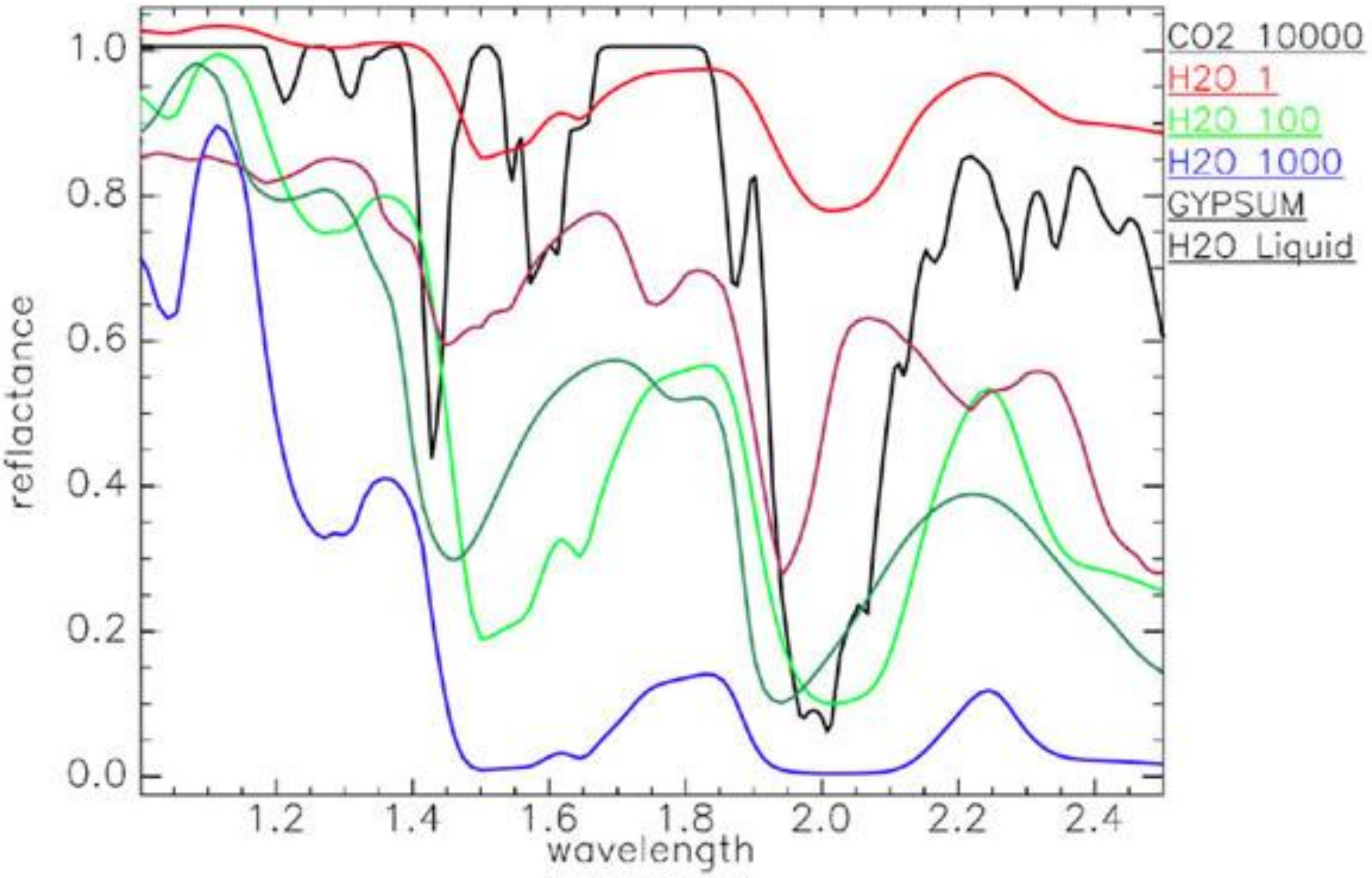

Fig. 13. Fraction of water ice spectra with 1 micron grain size from the total water ice content (grain size with 1 micron and 100 microns) as a function of time for the four units (see legend of Fig. 5). Dark core show the smaller grain size fraction. Grey ring and bright halo have intermediated grain size. Undisturbed terrains have the highest grain size. There is a general trend with increasing grain size from $L_S$=180 to 220 and then decreasing from $L_S$=220 to 240.

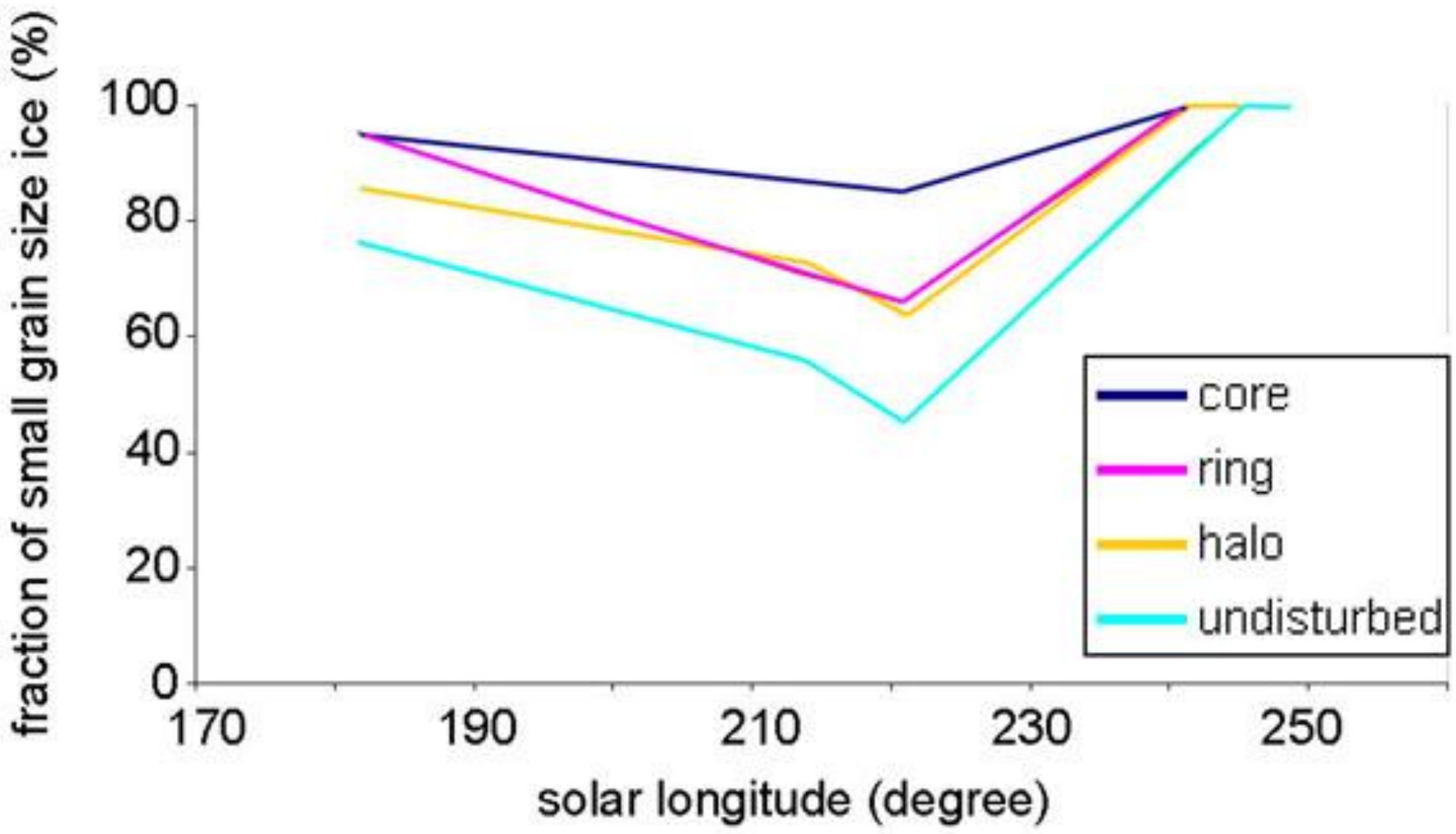

Fig. 14. Best synthetic spectra obtained by sparse linear mixture of reference database using FCLS algorithm versus observed spectra for the observation image no. 56CO dark core. Model 3 database include all spectra described in Fig. 12. In this case, liquid water is detected with an abundance of 0.04% and gypsum with an abundance of 1.57 % (rms=0.0056). Model 2 represents the same inversion with liquid water (1.46%) but without gypsum (rms=0.0058). Model 1 represents the same inversion without gypsum, nor liquid water (rms=0.0059). CRISM signal to noise ratio is 300 and thus errors bars cannot be seen in this figure. The rms is slightly decreasing from model 1 to model 3, indicating the possible presence of liquid water or gypsum but non-linear radiative transfer inversion has to confirm this trend.

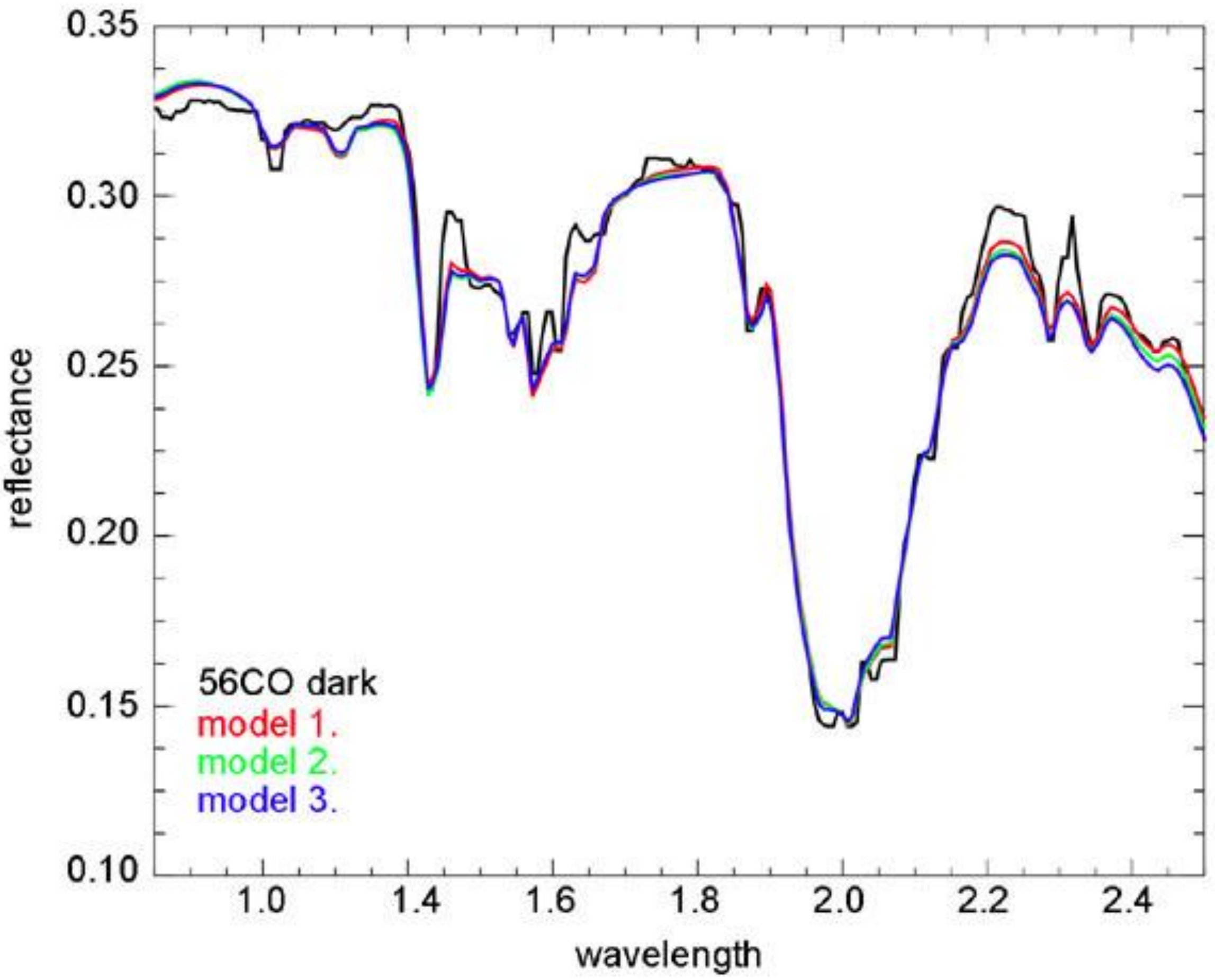

Fig. 15. The 4 first spectra at the top correspond to reference spectra (see Fig. 11) for $CO_2$ ice, water ice, liquid water and gypsum. Below ratio between undisturbed ice terrain divided by dark core for seven CRISM observations are visible: 43A2 ($L_S$=181.7°), 52BC ($L_S$=213.7°), 56CO ($L_S$=221.0°), 5C94 ($L_S$=241.7°), 5E38 ($L_S$=245.5°), 5FF6 ($L_S$=248.7°) and 6516 ($L_S$=262.6°). None of the ratio is compatible neither with gypsum nor with liquid water but with decreasing water ice and $CO_2$ ice compounds, strongly indicating that both water ice and $CO_2$ ice are present at the ground. The abundances of liquid water and gypsum, estimated by FCLS (model 3) on these ratio spectra, are always 0%. These results suggest that neither liquid water nor gypsum are enriched in the scene and therefore are presumably absent or in very small quantities.

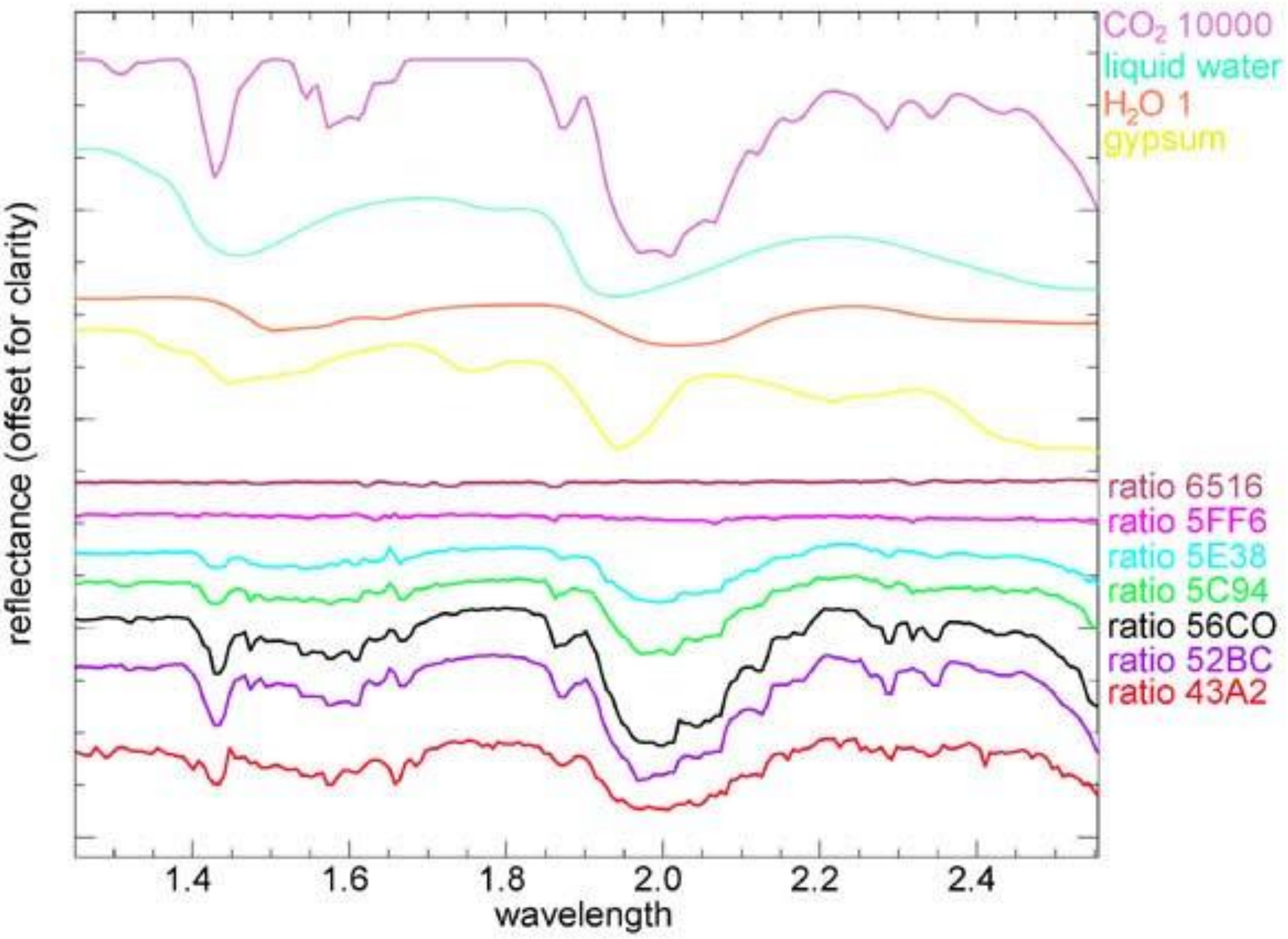

Fig. 16. Result of the model run on annual frost formation for $H_2O$ ice (black) and $CO_2$ ice (blue), plus the observed presence of any kind of ice at the target location from CRISM data (red). It shows as a function of $L_S$ (degree, horizontal axis) the cumulative thickness of $CO_2$ and $H_2O$ ice condensed on the ground, on the vertical axis the thickness of $H_2O$ and $CO_2$ are present using different scales, for $CO_2$ up to 40 cm and $H_2O$ up to 80 μm thickness. Diamonds, triangles and crosses correspond to 3 hypotheses for model parameters related to ground properties (see text for details).

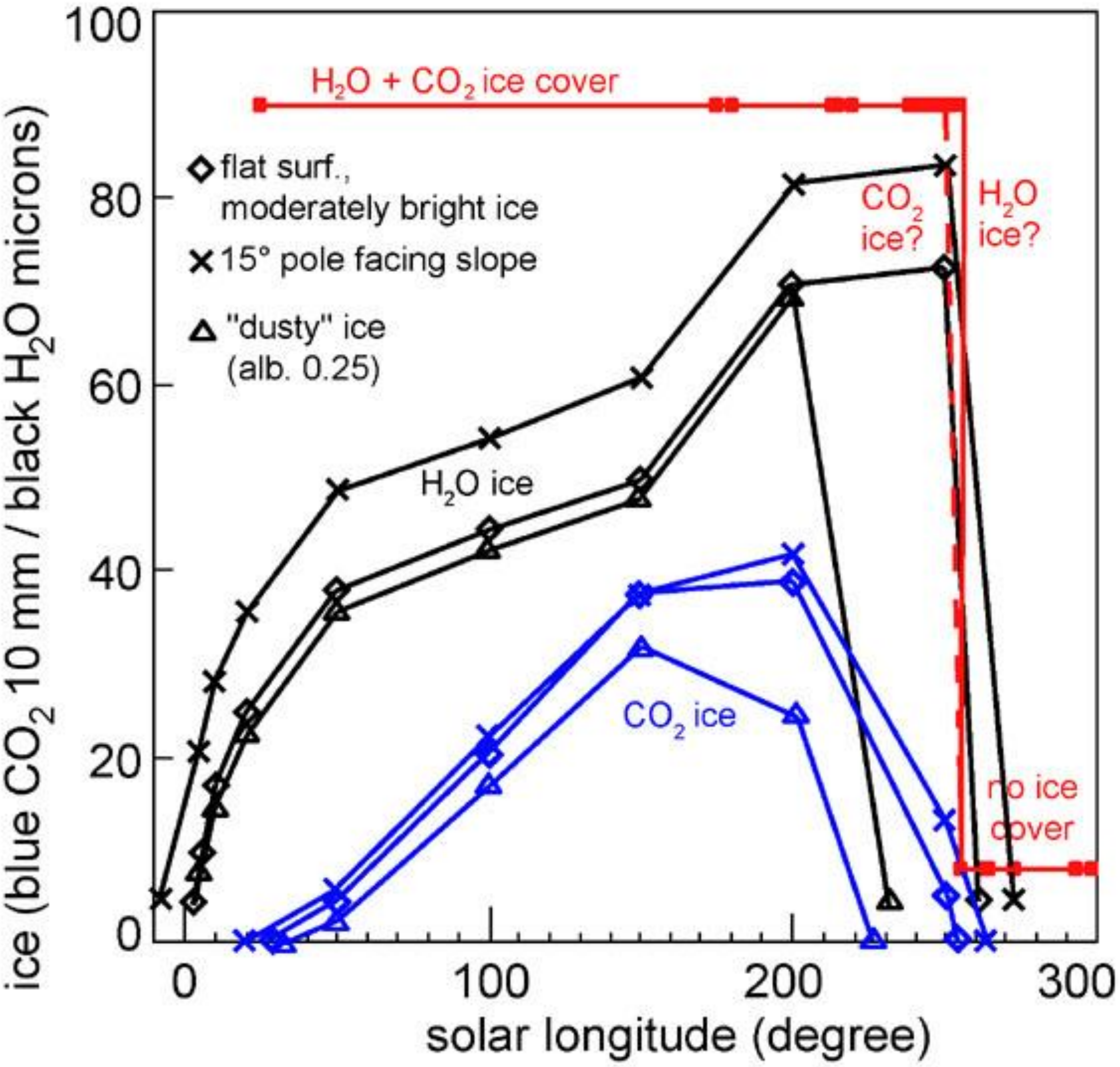